\documentclass[12pt,letterpaper]{article}
\pdfoutput=1
\usepackage[utf8]{inputenc}
\usepackage{amsmath}
\usepackage{slashed}
\usepackage{amssymb}
\usepackage{amsthm}
\usepackage{xcolor}
\definecolor{nicered}{rgb}{0.7,0.1,0.1}
\definecolor{nicegreen}{rgb}{0.1,0.5,0.1}
\usepackage[colorlinks=true,citecolor=nicegreen,linkcolor=nicered]{hyperref}
\usepackage{graphicx}
\graphicspath{{figures/}}
\usepackage{siunitx} 
\usepackage{cancel}
\usepackage{cite}
\usepackage{soul}
\usepackage{multirow}

\title{
Phenomenological consistency of the singlet-triplet scotogenic model
}

\author{ 
    Diego Restrepo\footnote{\href{mailto:restrepo@udea.edu.co}{restrepo@udea.edu.co}},
       Andrés Rivera\footnote{\href{mailto:afelipe.rivera@udea.edu.co}{afelipe.rivera@udea.edu.co}}
    \\
\textit{\small  Instituto de F\'{i}sica, Universidad de Antioquia,} \\
\textit{\small  Calle 70 No. 52-21, Medell\'{i}n, Colombia}
}
\date{\small \today}
\begin{document}

\maketitle
\begin{abstract}
We perform a complete analysis of the consistency of the singlet-triplet scotogenic model, where both dark matter and neutrino masses can be explained.
We determine the parameter space that yields the proper thermal relic density been in agreement with neutrino physics, lepton flavor violation, direct and indirect dark matter searches.
In particular, we calculate the dark matter annihilation into two photons, finding that the corresponding cross-section is below the present bounds reported by the Fermi-LAT and H.E.S.S. collaborations. 
We also determine the spin-dependent cross-section for dark matter elastic scattering with nucleons at one-loop level, finding that the next generation of experiments as LZ and DARWIN could test a small region of the parameter space of the model.
\end{abstract}

\section{Introduction}

There is solid evidence that supports the existence of Dark Matter (DM)~\cite{Zwicky:1937zza,Rubin:1970zza,Rubin:1980zd,clowebradacgonzalez2006,Refregier:2003ct,Tyson:1998vp}. Currently, it is well established that DM makes up about $27\%$ of the energy density of the Universe~\cite{Aghanim:2018eyx}. However, its nature and properties remain an open puzzle.
Additionally to the DM problem, the Standard Model (SM) has other open issue related with the fact that neutrinos are massive, which has been confirmed by neutrino oscillation experiments~\cite{deSalas:2017kay}.

In this article, we study these two puzzles within the singlet-triplet scotogenic model~\cite{Hirsch:2013ola}, which combines the scotogenic proposal~\cite{Ma:2006km} with the triplet fermion DM model~\cite{Ma:2008cu}. 
This framework is dubbed as the singlet-triplet fermion dark matter model or STFDM model for short. 
Their phenomenology was study in great detail in Refs.~\cite{Rocha-Moran:2016enp,Diaz:2016udz,Merle:2016scw}. However, in Ref.~\cite{Rocha-Moran:2016enp} authors studied the LFV observables taken into account the neutrino physics, but without the relic abundance of DM, in Ref.~\cite{Diaz:2016udz} authors studied the collider signals associated to the scalar sector, no the fermion sector, and in Ref.~\cite{Merle:2016scw} authors focus their attention in to study the consistency of the discrete symmetries of the model to high energies. 

The STFDM model has a rich phenomenology, with signals for WIMP-nucleons recoils that can be tested in future experiments like XENON1T~\cite{Aprile:2018dbl}, LZ~\cite{Akerib:2018lyp} and DARWIN~\cite{Aalbers:2016jon}.
Remarkably, the original proposal~\cite{Hirsch:2013ola} features spin-independent (SI) interactions of DM with nucleons and it is blind to spin-dependent (SD) interactions, since DM does not interact with the $Z$ gauge boson at tree-level. However, this observable can be generated at one-loop level as we will show later.
Other interesting aspect of the STFDM model is that it has lepton flavor violation (LFV) processes, such as $l_{\alpha}\to\,l_{\beta}\gamma$, 3-body decays as $\mu\to 3 e$, and $\mu - e$ conversion in nuclei that imply strong constraints on the parameter space~\cite{Rocha-Moran:2016enp}. 
Also, it was shown that the STFDM model is consistent to high energies. Specifically, the $\mathbb{Z}_2$ symmetry that stabilizes the DM particle and ensures the radiative seesaw mechanism for neutrino masses is preserved in the evolution of the renormalization group equation thanks to the presence of the scalar content of the model~\cite{Merle:2016scw}.

In this work, we study the full consistency of the STFDM model by performing a comparative analysis of a variety of observables. We find the parameter space that fulfills the relic density~\cite{Aghanim:2018eyx}, the neutrino physics parameters~\cite{deSalas:2017kay}, the LFV observables, and the direct-indirect searches of DM. 
Then, we explore the observables at one-loop level as the DM annihilation into two photons ($\text{DM}\,\text{DM}\to\gamma\gamma$) and the SD cross-section for elastic scattering with nucleons with the aim of obtain new DM observables.
Finally, we present the future prospects for fermionic DM in the STFDM model.

This paper is organized as follows. In Sec.~\ref{sec:model}, we introduce the STFDM model, in Sec.~\ref{sec:full-scan}, we present a broad scan of the parameter space that is consistent with DM, neutrino physics and the theoretical constraints, taking into account the perturbation character of the theory and the \textit{co-positivity} of the scalar potential. 
In Sec.~\ref{sec:indirect-direct-detection}, we analyze the direct and indirect detection status and its future prospects. 
In Sec.~\ref{sec:LFV}, we analyze the more restricted LFV processes.
In Sec.~\ref{sec:collider}, we do a final check using collider phenomenology for the fermionic production of DM.
In Sec.~\ref{sec:1-loop-processes}, we compute the new observables at one-loop level.
Specifically, we compute the SD cross-section and the DM annihilation into two photons. As far we know, those two expressions are reported for the first time.
Finally, in Sec.~\ref{sec:conclusions}, we summarize our results and present our outlook.

\section{The STFDM model}
\label{sec:model}

The STFDM model extends the gauge symmetry of the SM with a new discrete $\mathbb{Z}_2$ symmetry that stabilize the DM particle. In addition to the SM particle content, all even under the $\mathbb{Z}_2$ symmetry, the STFDM model is extended with a scalar doublet $\eta$, a real scalar triplet $\Omega$, and two fermions with zero hypercharge: a singlet $N$ and a triplet $\Sigma$. Their charge assignment is shown in Table~\ref{tab:new-particle-content}. 
In this work, we follow the notation given in~\cite{Merle:2016scw, Rocha-Moran:2016enp}. Explicitly, the new fields are,
\begin{align}
\eta = 
\begin{pmatrix}
\eta^+ \\
\eta^0
\end{pmatrix}
=\begin{pmatrix}
\eta^+ \\
\dfrac{1}{\sqrt{2}}\left(\eta^R+i\eta^I\right)
\end{pmatrix}, \;\hspace{0.3 cm}
\Omega = 
\begin{pmatrix}
\dfrac{\Omega^0}{\sqrt{2}} & \Omega^+ \\
\Omega^- & -\dfrac{\Omega^0}{\sqrt{2}}
\end{pmatrix}, \;\hspace{0.3 cm}
N\,, \;\hspace{0.3 cm}
\Sigma = 
\begin{pmatrix}
\dfrac{\Sigma^0}{\sqrt{2}} & \Sigma^+ \\
\Sigma^- & -\dfrac{\Sigma^0}{\sqrt{2}}
\end{pmatrix}\,.
\end{align}

\begin{table}
\centering
\begin{tabular}{|c|c|c|c|c|}
\hline
 \multicolumn{1}{|c}{   }  &
 \multicolumn{2}{|c}{ Scalars  } &
  \multicolumn{2}{|c|}{ Fermions  }  \\
 \hline 
Particle & $\eta$ & $\Omega$ & $N$ & $\Sigma$\\
\hline
SU(2)$_L$ &2 &3 &1 &3\\
U(1)$_Y$ &1/2 &0 &0 &0\\
$\mathbb{Z}_2$ & - &+ &- &-\\
\hline
\end{tabular}
\caption{New particle content and charges under the SU(2)$_L\times
$U(1)$_Y\times\mathbb{Z}_2$ group.}
\label{tab:new-particle-content}
\end{table}

The most general and invariant Yukawa Lagrangian is given by
\begin{align}
\label{eq:full-lagrangian}
\mathcal{L}\, = &\, Y_e^{\alpha\beta}\bar{L}_{\alpha}\,\phi\, e_{\beta}
+  Y_N^{\alpha}\bar{L}_{\alpha}\,\tilde{\eta}\, N 
+  Y_{\Sigma}^{\alpha}\bar{L}_{\alpha}\,\tilde{\eta}\, \Sigma 
+  Y_{\Omega}\overline{\Sigma}\,\Omega\, N\nonumber \\
+& \dfrac{1}{2}M_{\Sigma}\overline{\Sigma^c}\,\Sigma
+ \dfrac{1}{2}M_{N}\overline{N^c}N + h.c.\,,
\end{align}
where $L$ and $e$ are the SM fermions, $\alpha,\beta=1,2,3$, $\phi$ is the SM Higgs doublet and $\tilde{\eta}=i\sigma_2\eta^*$. On the other hand, the scalar potential of the STFDM model is given by
\begin{align}
\label{eq:scalar-potential}
V(\phi,\eta, \Omega) = &
-m_{\phi}^2 \phi^{\dagger}\phi + m_{\eta}^2 \eta^{\dagger}\eta
+\dfrac{1}{2}\lambda_1 (\phi^{\dagger}\phi)^2
+\dfrac{1}{2}\lambda_2 (\eta^{\dagger}\eta)^2
+\lambda_3 (\phi^{\dagger}\phi)(\eta^{\dagger}\eta)
+\lambda_4 (\phi^{\dagger}\eta)(\eta^{\dagger}\phi) \nonumber \\
+& \dfrac{\lambda_5}{2}\left[(\phi^{\dagger}\eta) + \text{h.c.}\right]
-\dfrac{m_{\Omega}^2}{2}\,\Omega^{\dagger}\Omega
+\dfrac{1}{2}\lambda_1^{\Omega} (\phi^{\dagger}\phi)(\Omega^{\dagger}\Omega)
+\dfrac{1}{4}\lambda_2^{\Omega} (\Omega^{\dagger}\Omega)^2
+\dfrac{1}{2}\lambda^{\eta} (\eta^{\dagger}\eta)(\Omega^{\dagger}\Omega)\nonumber \\
+& \mu_1\,\phi^{\dagger}\,\Omega\,\phi + \mu_2\,\eta^{\dagger}\,\Omega\,\eta\,.
\end{align}
This potential is subject to some theoretical constraints. First, we demand that all couplings $\lambda$ need to be $\leq 1$ to ensure the perturbativity of the theory and because they impact directly to the LFV processes as we will show latter. 
Second, we demand the stability of the potential (bounded from below). In this case, it has been shown that for $\lambda_4+|\lambda_5| \geq 0$, the \textit{co-positivity} of the potential is guaranteed if~\cite{Kannike:2012pe,Merle:2016scw};
\begin{align}
\label{eq:co-positivity}
&\lambda_1 \geq 0,\hspace{0.7 cm}  
\lambda_2 \geq 0,\hspace{0.7 cm} 
\lambda_2^{\Omega} \geq 0, \hspace{0.7 cm} 
\lambda_3+\sqrt{\lambda_1\lambda_2}\geq 0,\nonumber \\
&\lambda_3+\lambda_4-|\lambda_5|+\sqrt{\lambda_1\lambda_2}\geq 0,\hspace{0.7 cm} 
\lambda_1^{\Omega}+\sqrt{2\lambda_1\lambda_2^{\Omega}}\geq 0,\hspace{0.7 cm} 
\lambda^{\eta}+\sqrt{2\lambda_2\lambda_2^{\Omega}}\geq 0,\nonumber\\
&\sqrt{2\lambda_1\lambda_2\lambda_2^{\Omega}}+\lambda_3\sqrt{2\lambda_2^{\Omega}}
+\lambda_1^{\Omega}\sqrt{\lambda_2}
+\lambda^{\eta}\sqrt{\lambda_1} \,+\nonumber\\
&\sqrt{\left(\lambda_3+\sqrt{\lambda_1\lambda_2}\right)
\left(\lambda_1^{\Omega}+\sqrt{2\lambda_1\lambda_2^{\Omega}}\right)
\left(\lambda^{\eta}+\sqrt{2\lambda_2\lambda_2^{\Omega}}\right)}\geq 0\,,
\end{align}
where we should replace $\lambda_3$ by $(\lambda_3+\lambda_4-|\lambda_5|)$ in the last inequality in case that $\lambda_4+|\lambda_5|<0$.

The symmetry breaking in the STFDM model is such that 
\begin{align}
\label{eq:VEVs}
\langle \phi^0\rangle = \dfrac{v_{\phi}}{2},\hspace{0.5 cm} \langle \Omega^0 \rangle = v_{\Omega},\hspace{0.5 cm} \langle \eta^0 \rangle=0\,,
\end{align}
where the vacuum expectation values (VEVs) are themselves determinated by the tadpoles equations
\begin{align}
\label{eq:tadpole-phi}
t_{\phi}&=\dfrac{\partial V}{\partial v_{\phi}} = 
-m_{\phi}^2\,v_{\phi}+\dfrac{1}{2}\lambda_1v_{\phi}^3+ \dfrac{1}{2}\lambda_1^{\Omega}v_{\phi}v_{\Omega}^2-\dfrac{1}{\sqrt{2}}v_{\phi}v_{\Omega}\,\mu_1 =0 \,, \\
\label{eq:tadpole-Omega}
t_{\Omega}&=\dfrac{\partial V}{\partial v_{\Omega}} = 
-m_{\Omega}^2\,v_{\Omega}+\lambda_2^{\Omega}v_{\Omega}^3+ \dfrac{1}{2}\lambda_1^{\Omega}v_{\phi}^2v_{\Omega}-\dfrac{1}{\sqrt{2}}v_{\phi}^2\,\mu_1 =0\,.
\end{align}

In this frame, the $Z$ gauge boson receives a new contribution to its mass. The $W$ and $Z$ gauge bosons masses are given by

\begin{align}
\label{eq:mW-mZ}
m_W^2 &= \dfrac{1}{4}g^2\left(v_{\phi}^2+4v_{\Omega}^2\right)\,,\hspace{0.5 cm} 
m_Z^2=\dfrac{1}{4}\left(g^2+g^{'2}\right)v_{\phi}^2\,.
\end{align}
In particular, the $W$ boson mass is strongly constrained by the value of the triplet VEV, we demand that $v_{\Omega}< 5$~GeV~\cite{Tanabashi:2018oca}.

\subsection{$\mathbb{Z}_2$-even and $\mathbb{Z}_2$-odd spectrum}

The scalar spectrum is divided in two parts: The $\mathbb{Z}_2$-even scalars $\phi^0$, $\Omega^0$, $\Omega^{\pm}$, $\phi^{\pm}$ and the $\mathbb{Z}_2$-odd scalars $\eta^0$, $\eta^{\pm}$, where $\eta^0$ is a good DM candidate widely studied in the literature\cite{Deshpande:1977rw,Barbieri:2006dq,LopezHonorez:2006gr,Honorez:2010re,Garcia-Cely:2015khw,Queiroz:2015utg,Diaz:2016udz}.
In this frame, the neutral scalars $\phi^0$ and $\Omega^0$ are mixed by a $2\times 2$ mass matrix, which can be parametrized with the angle $\beta$, such that
\begin{align}
\label{eq:beta-mixing}
\begin{pmatrix}
h_1 \\ h_2 
\end{pmatrix}=
\begin{pmatrix}
\cos\beta & \sin\beta \\
-\sin\beta & \cos\beta
\end{pmatrix}
\begin{pmatrix}
\phi^0 \\ \Omega^0 
\end{pmatrix}\,,
\end{align} 
where
\begin{equation}
\tan(2\beta) = \frac{4 v_{\Omega } v_{\phi } \left(\sqrt{2} \mu _1-2 \lambda _1^{\Omega } v_{\Omega }\right)}{8 \lambda _2^{\Omega } v_{\Omega
   }^3-4 \lambda _1 v_{\Omega } v_{\phi }^2+\sqrt{2} \mu _1 v_{\phi }^2}\,.
\end{equation}
The lightest $\mathbb{Z}_2$-even scalar $h_1$ will be identified with the $125$~GeV scalar of the SM and the heavier one will remain as a new scalar Higgs boson present in this theory.  
In the same way, the charged scalars $\phi^{\pm}$ and $\Omega^{\pm}$ are also mixed by a $2\times 2$ mass matrix,
\begin{align}
\label{eq:delta-mixing}
\begin{pmatrix}
h_1^{\pm} \\ h_2^{\pm} 
\end{pmatrix}=
\begin{pmatrix}
\cos\delta & \sin\delta \\
-\sin\delta & \cos\delta
\end{pmatrix}
\begin{pmatrix}
\phi^{\pm} \\ \Omega^{\pm} 
\end{pmatrix}\,,
\end{align}
with
\begin{equation}
\tan(2\delta) = -\frac{4 v_{\Omega } v_{\phi }}{v_{\phi }^2-4 v_{\Omega }^2}\,.
\end{equation}
The lightest $h_1^{\pm}$ charged scalar needs to be identified with the Goldstone boson  which is the longitudinal component of the $W$ boson. The other field is identified as a new charged scalar present in this theory.  
In addition, the masses of the $\mathbb{Z}_2$-odd scalars $\eta^{\pm}$ and $\eta^0$ are given by
\begin{align}
\label{eq:masa-etapm}
m_{\eta^\pm}^2 &= m_{\eta}^2 + \dfrac{1}{2}\lambda_3v_{\phi}^2 
+ \dfrac{1}{2}\lambda^{\eta}v_{\Omega}^2 + \dfrac{1}{\sqrt{2}}v_{\Omega}\,\mu_2\,, \\
\label{eq:masa-etaR}
m_{\eta^R}^2 &= m_{\eta}^2 + \dfrac{1}{2}\left(\lambda_3+\lambda_4+\lambda_5\right) v_{\phi}^2
+ \dfrac{1}{2}\lambda^{\eta}v_{\Omega}^2 - \dfrac{1}{\sqrt{2}}v_{\Omega}\,\mu_2\,, \\
\label{eq:masa-etapI}
m_{\eta^I}^2 &= m_{\eta}^2 + \dfrac{1}{2}\left(\lambda_3+\lambda_4-\lambda_5\right) v_{\phi}^2
+ \dfrac{1}{2}\lambda^{\eta}v_{\Omega}^2 - \dfrac{1}{\sqrt{2}}v_{\Omega}\,\mu_2\,.
\end{align}

On the other hand, the new fermion spectrum consists of two neutral fermions $\chi_i^0$, of which the lightest one can be the DM particle, and one charged fermion $\chi^{\pm}$~\footnote{The mass of the $\chi^{\pm}$ particle at tree-level is given by $M_{\Sigma}$, however, it is known that there is a mass gap between the $\Sigma^0$ and $\Sigma^{\pm}$ in the pure triplet fermion model which is approximately given by the mass of the neutral Pion $\pi^0$~\cite{Cirelli:2005uq, Choubey:2017yyn}.}. 
Explicitly, the $\mathbb{Z}_2$-odd fields $\Sigma^0$ and $N$ are mixed by the Yukawa coupling $Y_{\Omega}$ of Eq.~\eqref{eq:full-lagrangian} and a non-zero VEV $v_{\Omega}$. The Majorana mass matrix in the basis $(\Sigma^0, N)$, is given by
\begin{align}
\label{eq:M-chi-matrix}
M_{\chi}=
\begin{pmatrix}
M_{\Sigma} & Y_{\Omega}v_{\Omega} \\
Y_{\Omega}v_{\Omega} & M_N
\end{pmatrix}\,,
\end{align}
which is diagonalized by a $2\times2$ matrix $V(\alpha)$, 
\begin{align}
\label{eq:M-chi-rotation}
\begin{pmatrix}
\chi_1^0 \\
\chi_2^0
\end{pmatrix}=
V(\alpha)
\begin{pmatrix}
\Sigma^0 \\
N
\end{pmatrix}
=
\begin{pmatrix}
\cos\alpha & \sin\alpha \\
-\sin\alpha & \cos\alpha
\end{pmatrix}
\begin{pmatrix}
\Sigma^0 \\
N
\end{pmatrix}\,.
\end{align}
Therefore, the tree-level mass for the $\chi^{\pm}$ and the $\chi^0_i$ eigenstates are
\begin{align}
m_{\chi^{\pm}} &= M_\Sigma\,, \nonumber \\
m_{\chi_1^0} &= \dfrac{1}{2}\left(M_\Sigma + M_N - \sqrt{(M_\Sigma - M_N)^2+4(Y_{\Omega}v_{\Omega})^2} \right)\,, \nonumber \\
\label{eq:chi-masses}
m_{\chi_2^0} &= \dfrac{1}{2}\left(M_\Sigma + M_N + \sqrt{(M_\Sigma - M_N)^2+4(Y_{\Omega}v_{\Omega})^2} \right)\,,
\end{align}
and the mixing angle $\alpha$ fulfill the relation
\begin{align}
\label{eq:tan-alpha}
\tan(2\alpha)= \dfrac{2Y_{\Omega}v_\Omega}{M_{\Sigma}-M_N}\,.
\end{align}

\subsection{Dark matter candidates}
\label{sec:dark-matter}

The STFDM model could have scalar and fermionic candidates for DM particle. 
\begin{enumerate}
\item[i)] Regarding scalar DM, the lightest component of the neutral state $\eta^0$ is the DM candidate. 
This case has been studied extensively in the literature~\cite{Deshpande:1977rw,Barbieri:2006dq,LopezHonorez:2006gr,Honorez:2010re,Garcia-Cely:2015khw,Queiroz:2015utg,Diaz:2016udz} and it is known that its phenomenology is driven principally for gauge interactions which dominate the DM production in the early universe. 

\item[ii)] Regarding fermion DM, the lightest eigenvalue $\chi_1^0$ that comes from the mixing between the triplet component $\Sigma^0$ and the fermion singlet $N$ is the DM candidate. In this case, we have a interesting phenomenology that comes from the mixing between the singlet and the triplet fermion~\cite{Hirsch:2013ola, Rocha-Moran:2016enp, Merle:2016scw}. Even more, some important features of this DM candidate are based on its nature itself. When it is principally singlet ($\chi_1^0\approx N$), the DM phenomenology is dominated by the Yukawa interactions, principally driven by the $Y_N$ coupling of the Lagrangian~\eqref{eq:full-lagrangian}. It implies some direct relation with LFV observables and it is difficult to explain the relic abundance with Yukawa coupling to order $\mathcal{O}\lesssim 1$~\cite{Ibarra:2016dlb}. 
On the other hand, when the DM is mostly triplet ($\chi_1^0\approx \Sigma^0$), its phenomenology is driven by gauge interaction. The coannihilation between DM and $\chi^{\pm}$ is really important and there is not serious implications on LFV observables. 
Furthermore, it is known that in this regime the correct relic density is only reproduced when the DM mass is around $\sim 2.4$~GeV~\cite{Ma:2008cu,Choubey:2017yyn}. 
Now, with the singlet-triplet mixing, some very features arise, perhaps, the most attractive one is that the mixing itself give us the opportunity to have a DM particle in the GeV-TeV range.     
In this paper, we will focus in the fermion DM case, which is the lightest eigenvalue $\chi_1^0$.
\end{enumerate}

\subsection{Neutrino masses}
\label{sec:neutrino-masses}

\begin{figure}
\begin{center}
\includegraphics[scale=0.4]{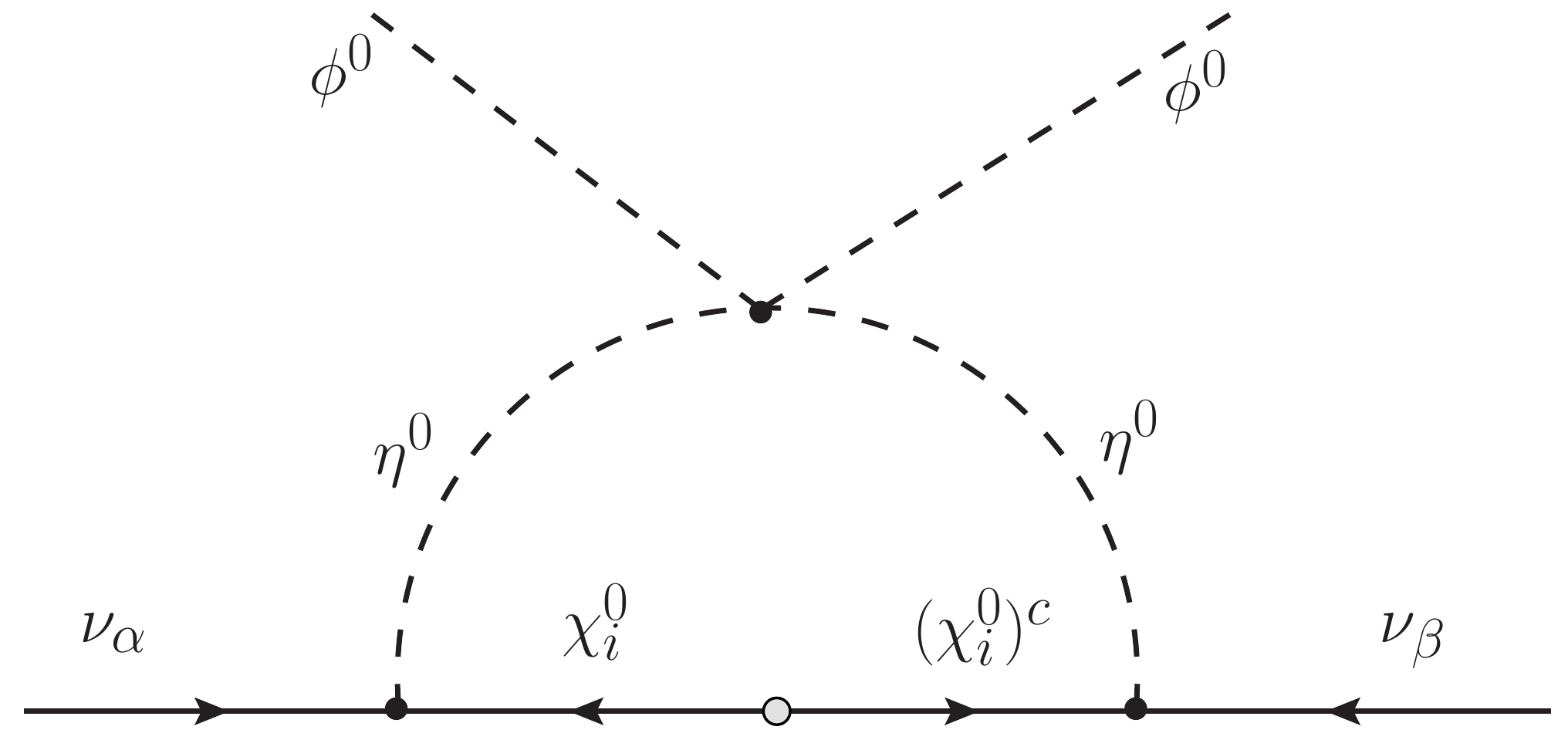}
\caption{One-loop diagram that generate neutrino masses. }
\label{fig:mass-diagram}
\end{center}
\end{figure}

In the STFDM model, the Majorana neutrino masses are generated at one-loop level as shown in Fig.~\ref{fig:mass-diagram}.
The neutrino mass matrix at one-loop level can be written as
\begin{align}
\label{eq:Mv-mass-matrix}
(\mathcal{M}_\nu)_{\alpha\beta} &= \sum_{i=1}^2\dfrac{h_{\alpha i}\,h_{\beta i}\,m_{\chi_i^0}}{2(4\pi)^2}
\left[
\dfrac{m_{\eta^R}^2\ln\left(\dfrac{m_{\chi_i^0}^2}{m_{\eta^R}^2}\right)}{m_{\chi_i^0}^2-m_{\eta^R}^2}
-\dfrac{m_{\eta^I}^2\ln\left(\dfrac{m_{\chi_i^0}^2}{m_{\eta^I}^2}\right)}{m_{\chi_i^0}^2-m_{\eta^I}^2}
\right] \nonumber\\
&= \sum_{i=1}^2h_{\alpha i}\,\Lambda_i\,(h^T)_{i\beta}
=(h\,\Lambda\,h^T)_{\alpha\beta}\,,
\end{align}
where $h$ and $\Lambda$ are matrices, respectively given by
\begin{align}
\label{eq:h-Lambda-matrices}
h &= \dfrac{1}{\sqrt{2}}
\begin{pmatrix}
Y_{\Sigma}^1 & \sqrt{2}\,Y_N^1 \\
Y_{\Sigma}^2 & \sqrt{2}\,Y_N^2 \\
Y_{\Sigma}^3 & \sqrt{2}\,Y_N^3
\end{pmatrix}
\cdot V^{T}(\alpha)\,,
\hspace{1 cm}
\Lambda=
\begin{pmatrix}
\Lambda_1 & 0 \\
0 & \Lambda_2
\end{pmatrix}\,,
\end{align}
with
\begin{align}
\label{eq:Lambdai-definition}
\Lambda_i = \dfrac{m_{\chi_i^0}}{2(4\pi)^2}
\left[
\dfrac{m_{\eta^R}^2\ln\left(\dfrac{m_{\chi_i^0}^2}{m_{\eta^R}^2}\right)}{m_{\chi_i^0}^2-m_{\eta^R}^2}
-\dfrac{m_{\eta^I}^2\ln\left(\dfrac{m_{\chi_i^0}^2}{m_{\eta^I}^2}\right)}{m_{\chi_i^0}^2-m_{\eta^I}^2}
\right] \,.
\end{align}
Note that in the limit of $m_{\eta^R}= m_{\eta^I}$ we have zero neutrino masses. This vanishing can be understood because according to Eqs.~\eqref{eq:masa-etaR} and~\eqref{eq:masa-etapI} it means that $\lambda_5$=0 and therefore in this model can be imposed a conserved lepton number. Even more, it can be shown that, in the limit where the $\chi_i^0$ eigenvalues are lighter than the other fields, we obtain a simple expression for the neutrino mass matrix in terms of $\lambda_5$~\cite{Hirsch:2013ola}, namely
\begin{align}
\label{eq:Mv-aproximations}
(\mathcal{M}_\nu)_{\alpha\beta} &\approx \sum_{i=1}^2\dfrac{h_{\alpha i}\,h_{\beta i}}{(4\pi)^2}
\dfrac{\lambda_5 v_{\phi}^2}{m_0^2}
m_{\chi_i^0}\,,
\end{align}  
where 
\begin{align}
\label{ec_mo-definition}
m_0^2 = m_{\eta}^2 + \dfrac{1}{2}\left(\lambda_3+\lambda_4\right) v_{\phi}^2
+ \dfrac{1}{2}\lambda^{\eta}v_{\Omega}^2 - \dfrac{1}{\sqrt{2}}v_{\Omega}\,\mu_2\;
 \Rightarrow \;
 m_{\eta^{I,R}}^2 =m_0^2\pm\lambda_5 v_{\phi}^2\,.
\end{align}
It is convenient express the Yukawa couplings $h_{\alpha i}$ in Eq.~\eqref{eq:Mv-mass-matrix} using the Casas-Ibarra parametrization~\cite{Casas:2001sr, Ibarra:2003up}. It turns out that
\begin{align}
h = U^*\sqrt{\widetilde{M}}\,R\,\sqrt{\Lambda}^{-1}\,, 
\end{align}
where $U$ is the PMNS (Pontecorvo-Maki-Nakagawa-Sakata) matrix, $\widetilde{M}=\text{diag}(m_1,m_2,m_3)$ with $m_i$ the neutrino physical masses, $\Lambda$ is given by Eq.~\eqref{eq:h-Lambda-matrices} and $R$ is a $3\times 2$ complex, arbitrary and orthogonal matrix, such that $R\,R^T=\mathbb{I}_{3\times3}$. The matrix $R$ is similar to that one found in the context of type-one seesaw with two generations of right-handed neutrinos, where we obtain one massless neutrino~\cite{Ibarra:2003up}. It depends on the neutrino hierarchy (NH: Normal hierarchy, IH: Inverse hierarchy),
\begin{align}
\label{eq:R-matrix}
R=\begin{pmatrix}
0 & 0 \\
\cos\gamma & \sin\gamma \\
-\sin\gamma & \cos\gamma
\end{pmatrix}\hspace{0.5 cm} &\text{for NH}\,,&\hspace{1.0 cm}
R=\begin{pmatrix}
\cos\gamma & \sin\gamma \\
-\sin\gamma & \cos\gamma \\
0 & 0
\end{pmatrix}\hspace{0.5 cm} &\text{for IH}& \nonumber \\
&m_1 \rightarrow 0&   &m_3 \rightarrow 0\,,& 
\end{align}
where $\gamma$ is in general a complex angle.

\section{Numerical results}
\label{sec:full-scan}

\begin{figure}
\begin{center}
\includegraphics[scale=0.43]{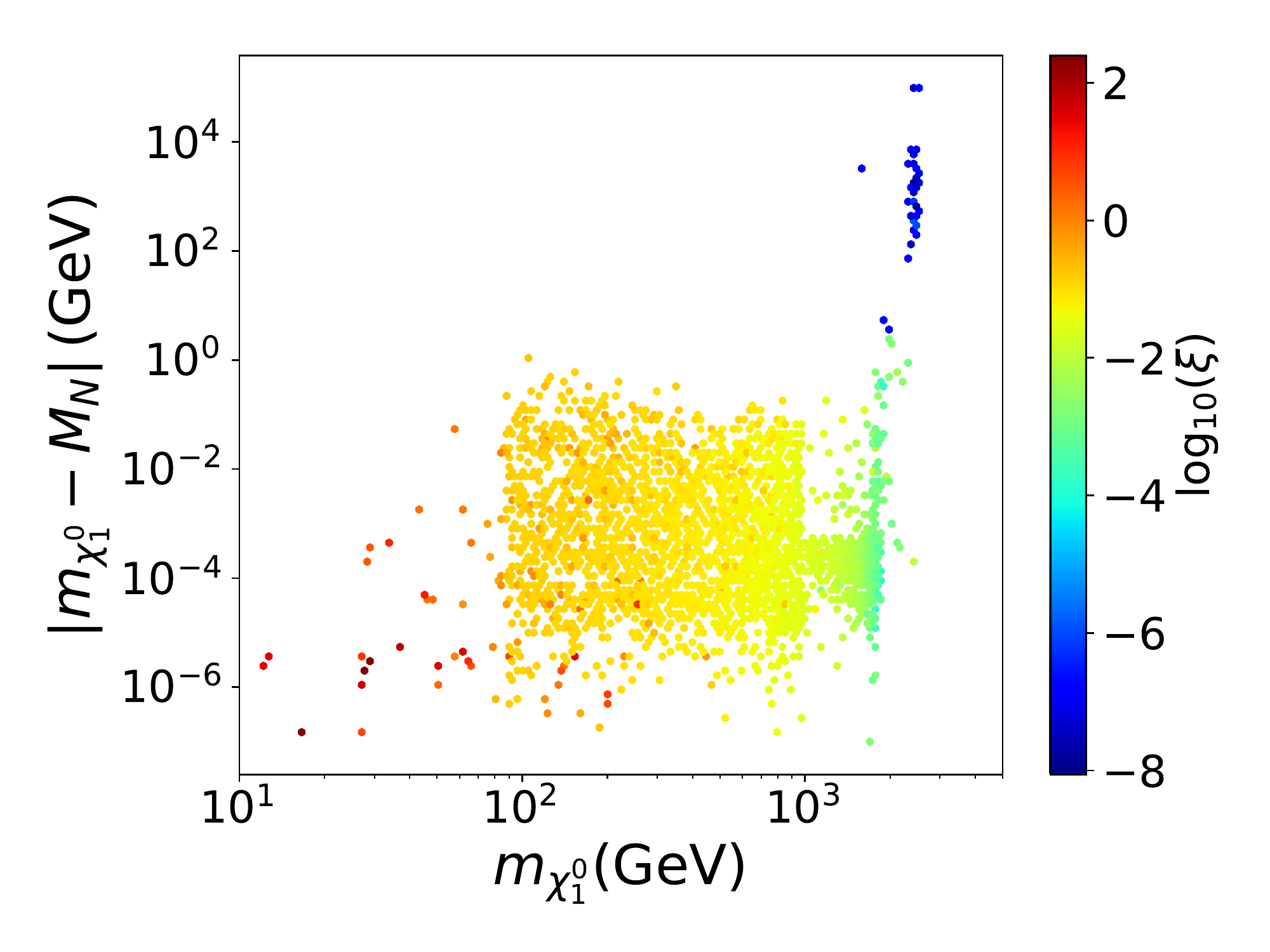}
\includegraphics[scale=0.43]{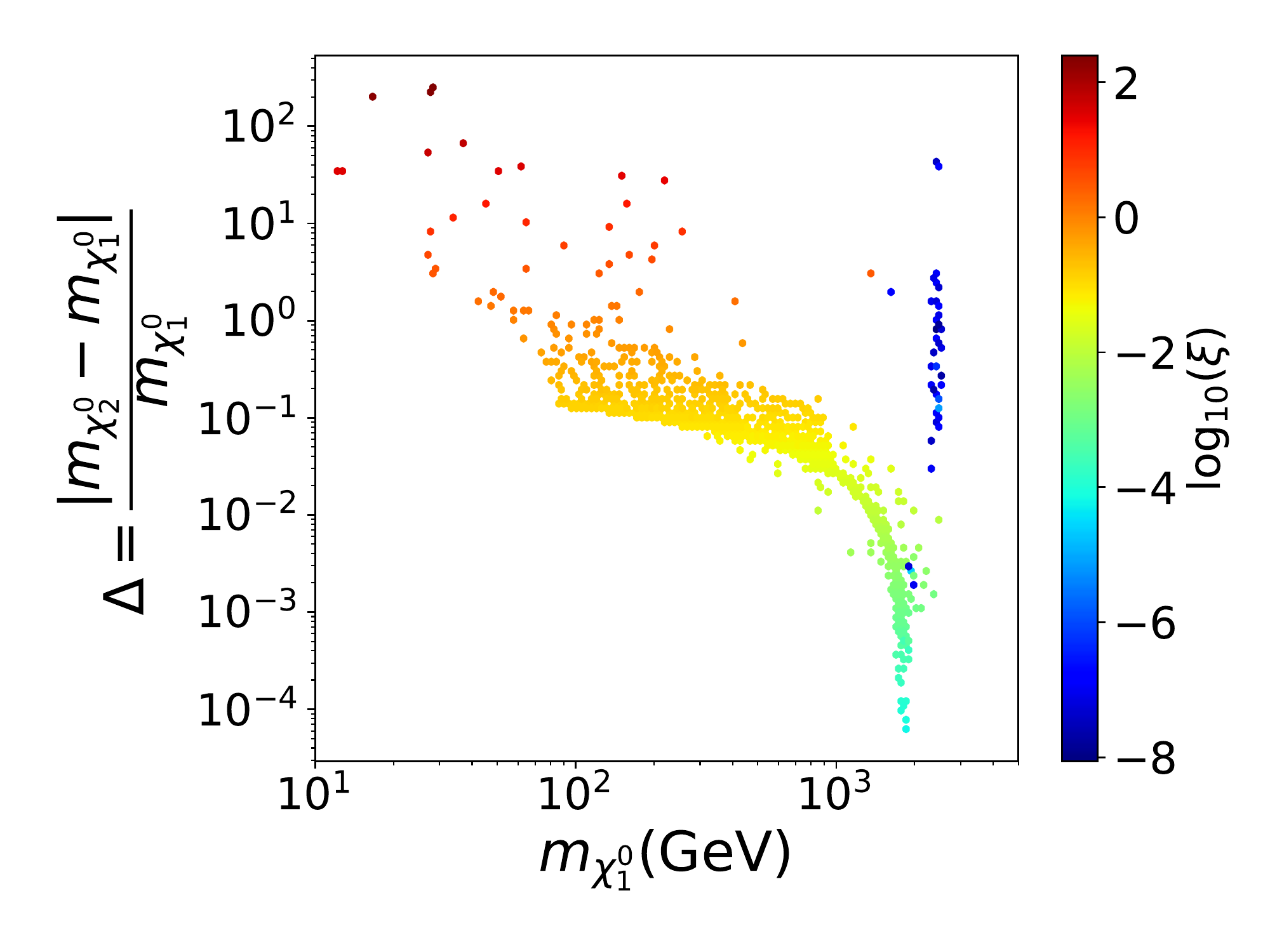}
\caption{Parameter space that is fully consistent with DM and neutrino physics in the STFDM model. In color, we show the variable $\xi$ define in Eq.~\eqref{eq:xi}. High values of $\xi$ is the limit for singlet fermion DM $\sim N$ and low values correspond to the limit for triplet fermion DM $\sim \Sigma^0$.}
\label{fig:MN-and-MTF}
\end{center}
\end{figure}

In order to study the DM phenomenology of the STFDM model, we have scanned the parameter space according to the ranges shown in Table~\ref{tab:scan-parameter}. We chose $m_{\eta}$ and $M_{\Sigma} > 100$~GeV in order to be conservative with LEP searches of charged particles~\cite{ALEPH:2005ab}. We also chose $v_{\Omega}<5$~GeV to be compatible with the $W$ gauge boson mass~\cite{Tanabashi:2018oca}. 
The remaining parameters were computed from this set. In particular, $m_{\Omega}$ was computed using Eq.~\eqref{eq:tadpole-Omega}, $\lambda_1$ and $m_{\phi}^2$ in the scalar potential were fixed by the tadpole Eq.~\eqref{eq:tadpole-phi} and the mass for the scalar of the SM ($m_{h_1}\approx 125$~GeV).
\begin{table}
\centering
\begin{tabular}{|c|l|}
\hline
Parameter & Range\\
\hline
$M_N$ &  $1-10^4$ (GeV) \\
$M_{\Sigma}$ &  $100-10^4$ (GeV) \\
$m_{\eta}$ &  $100-10^4$ (GeV) \\
$\mu_i$ & $1-10^{5}$ (GeV) \\
$|\lambda_{2,3,4}|$ , $|\lambda_i^{\Omega}|$ , $|\lambda^{\eta}|$, $|Y_{\Omega}|$ & $10^{-4} - 1$ \\
$|\lambda_5|$  & $10^{-10} - 1$ \\
$v_{\Omega}$ & $10^{-2}- 5$  (GeV)\\
\hline
\end{tabular}
\caption{Scanning parameter ranges.}
\label{tab:scan-parameter}
\end{table}
We did  a carefully random search where we imposed the theoretical constraints given by Eq.~\eqref{eq:co-positivity} and the correct Yukawa coupling $Y_{\Sigma}^i$, $Y_N^i$ that reproduced the neutrino oscillation parameters~\cite{Forero:2014bxa, deSalas:2017kay}. 
In order to do that, we followed the algorithm described in Sec.~\ref{sec:neutrino-masses}\footnote{We realized that neutrino hierarchy (IH, NH) does not play an important role in the analysis, for that reason we select randomly both hierarchies.}.
Also, we took into account the invisible decay of the Higgs boson~\cite{Choubey:2017yyn}, which demands an invisible branching fraction $<24\%$ at $95\%$ confidence level~\cite{Tanabashi:2018oca}.
We implemented the STFDM model in~\texttt{SARAH}~\cite{Staub:2008uz,Staub:2009bi,Staub:2010jh,Staub:2012pb,Staub:2013tta} couple to \texttt{SPheno}~\cite{Porod:2003um,Porod:2011nf} routines. 
Later, we used~\texttt{MicrOMEGAs 4.2.5}~\cite{Belanger:2006is} in order to compute the relic density and we only took the models that fulfill the current value $\Omega h^2 = (0.120 \pm 0.001)\; \text{to}\; 3\sigma$~\cite{Aghanim:2018eyx}.
We realized, although the mixture between the triplet fermion $\Sigma^0$ and the singlet fermion $N$ is important, the parameters space that is fully consistent with the DM framework and the neutrino physics prefers a singlet component in the low mass region.
This feature is shown in the left panel of Fig.~\ref{fig:MN-and-MTF} where we can see that $m_{\chi_1^0} \approx m_N $ for $m_{\chi_1^0}<2$~TeV. 
On the other hand, in the right panel of this figure, we show the parameter $\Delta=|m_{\chi_2^0}-m_{\chi_1^0}|/m_{\chi_1^0}$ that characterized the coannihilation processes in the STFDM model~\cite{Griest:1990kh}. 
We realized that coannihilation process between the singlet and the triplet fermion plays an important role and brings the relic density to its observed value for almost all the points with $80$~GeV $<m_{\chi_1^0}<2.4$~TeV. 
However, the points with  $m_{\chi_1^0}< 1$~TeV and $\Delta > 10$ will generate high LFV process that can rule out the STFDM model as we will show later. 
In general, we realized that the neutral fermion spectrum is almost degenerate for the majority of the points up to $2.4$~TeV. For masses larger than this value, the STFDM model recovers the known limit of the Minimal DM scenarios in which the DM particle is the triplet $\Sigma$. 
In order to have an intuition of the nature of the DM, we show in color the quantity
\begin{align}
\label{eq:xi}
\xi &=\dfrac{|M_\Sigma - m_{\chi^0_1}|}{m_{\chi^0_1}}
\end{align}
that was introduced in~\cite{Hirsch:2013ola}. Low values correspond to triplet DM and high values to singlet DM.

\subsection{The status of direct-indirect detection of dark matter }
\label{sec:indirect-direct-detection}

A tree-level, the STFDM model produces direct detection signals. In particular, it has recoils with nucleons that are SI  and it is blind to SD signals because it does not have a tree-level coupling between the DM and $Z$ gauge boson. 
\begin{figure}
\begin{center}
\includegraphics[scale=0.37]{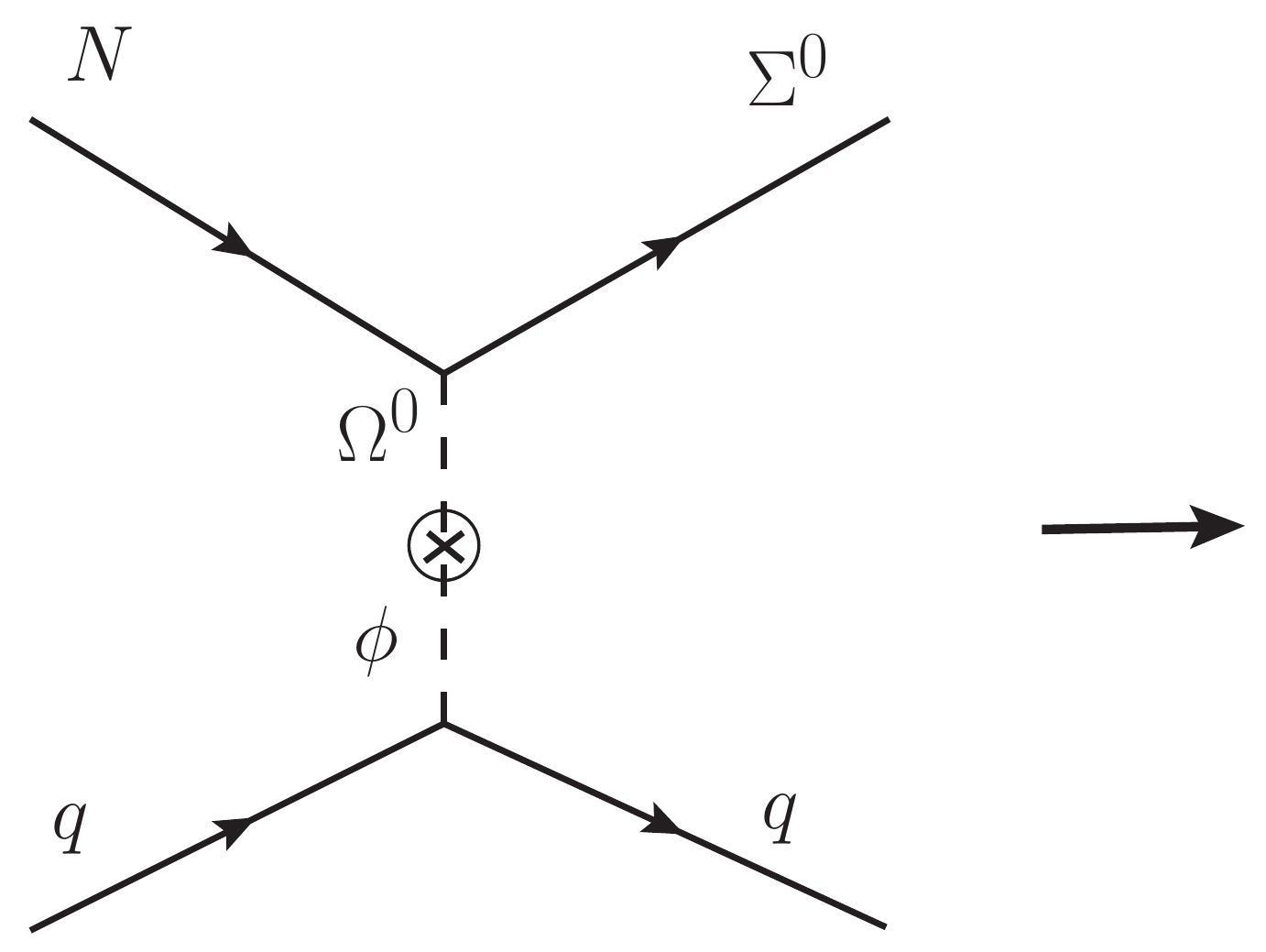}
\includegraphics[scale=0.37]{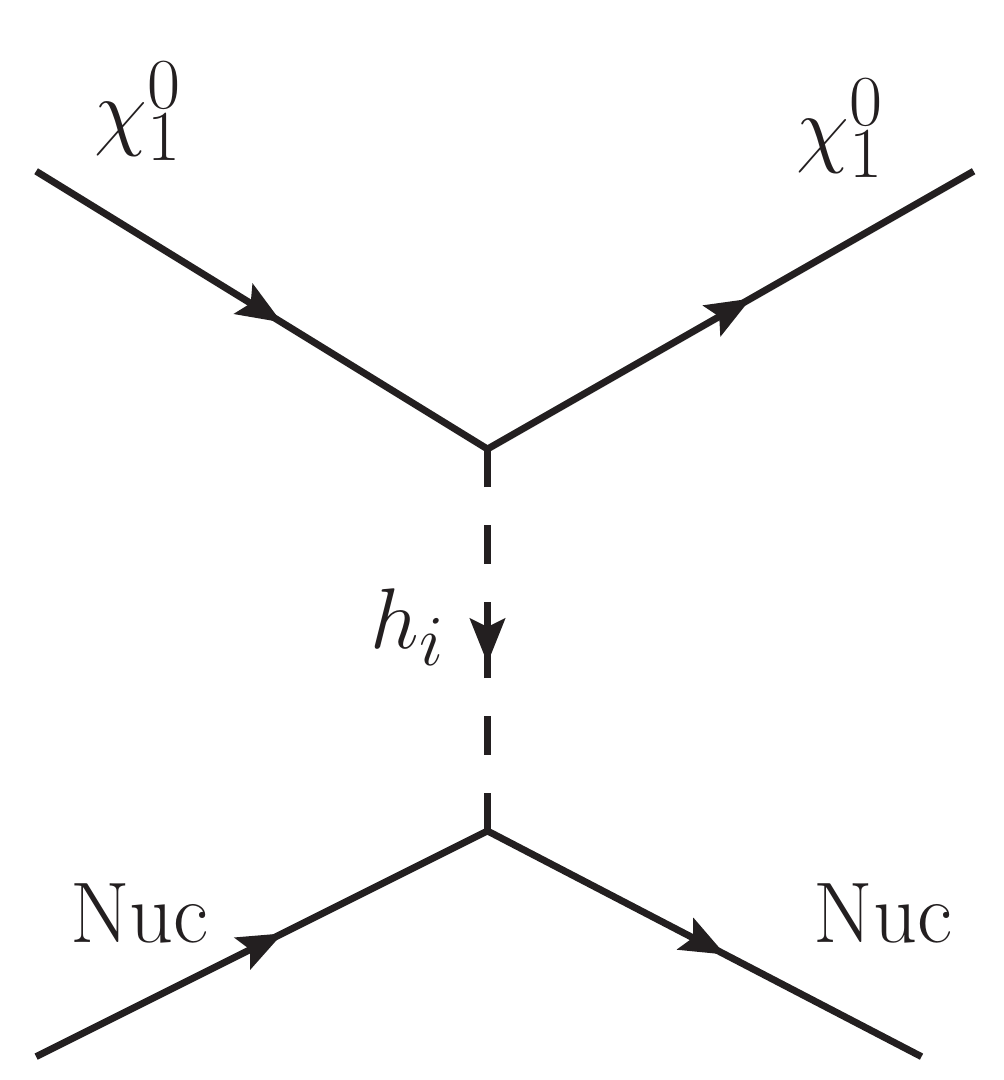}
\caption{SI process in the STFDM model. In the left, we show the process in the gauge basis. In the right, we show the process in the mass basis in order to emphasize that actually, we have two contributions coming from the Higgses $h_i$.}
\label{fig:SI-diagram}
\end{center}
\end{figure}
The SI scattering process is mediated by the two Higgses $h_i$ that result from the mixing between the scalars $\Omega^0$ and $\phi$. 
This process is shown in Fig.~\ref{fig:SI-diagram} and it is easily computed in the limit where the  Mandelstam variable $t$ is negligible. 
The scattering cross-section is given by
\begin{align}
\label{eq:sigma-SI}
\sigma_{SI}\approx\dfrac{\mu_{\text{red}}^2}{\pi}\left[\dfrac{M_{\text{Nuc}}f_N}{v}\dfrac{Y_{\Omega}\sin(2\alpha)\sin(2\beta)}{2}\left(\frac{1}{m_{h_2}^2}-\frac{1}{m_{h_1}^2}\right)\right]^2\,,
\end{align}
where, $M_{\text{Nuc}}$ is the nucleon mass, $f_N\approx 0.3$ is the nucleon form factor, $\mu_{\text{red}}= m_{\chi^0_1}M_{\text{Nuc}}/(m_{\chi^0_1}+M_{\text{Nuc}})$ is the reduced mass of the system, and  $m_{h_i}$ is the mass of the Higgses $h_i$.

\begin{figure}
\begin{center}
\includegraphics[scale=0.43]{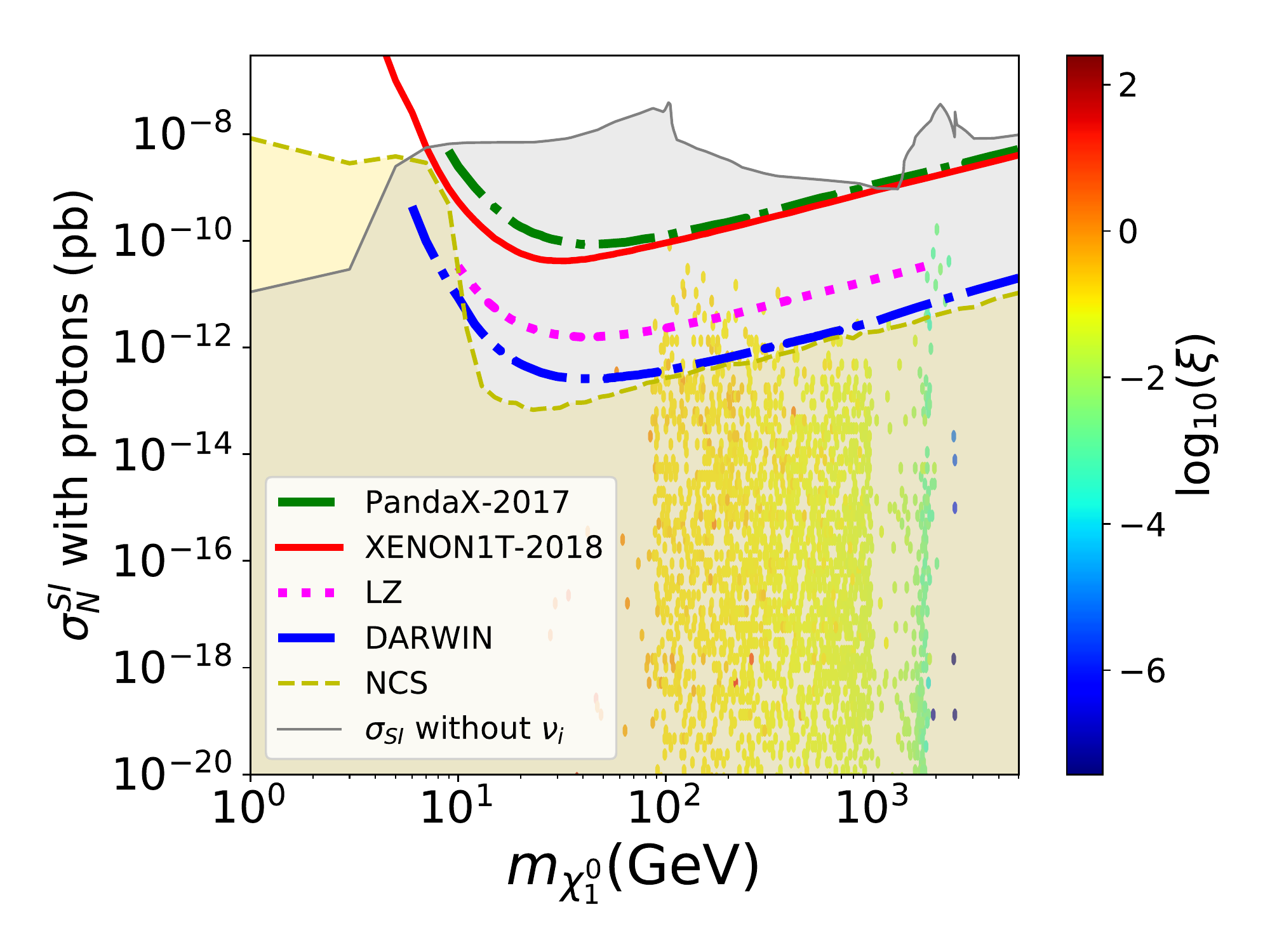}
\includegraphics[scale=0.43]{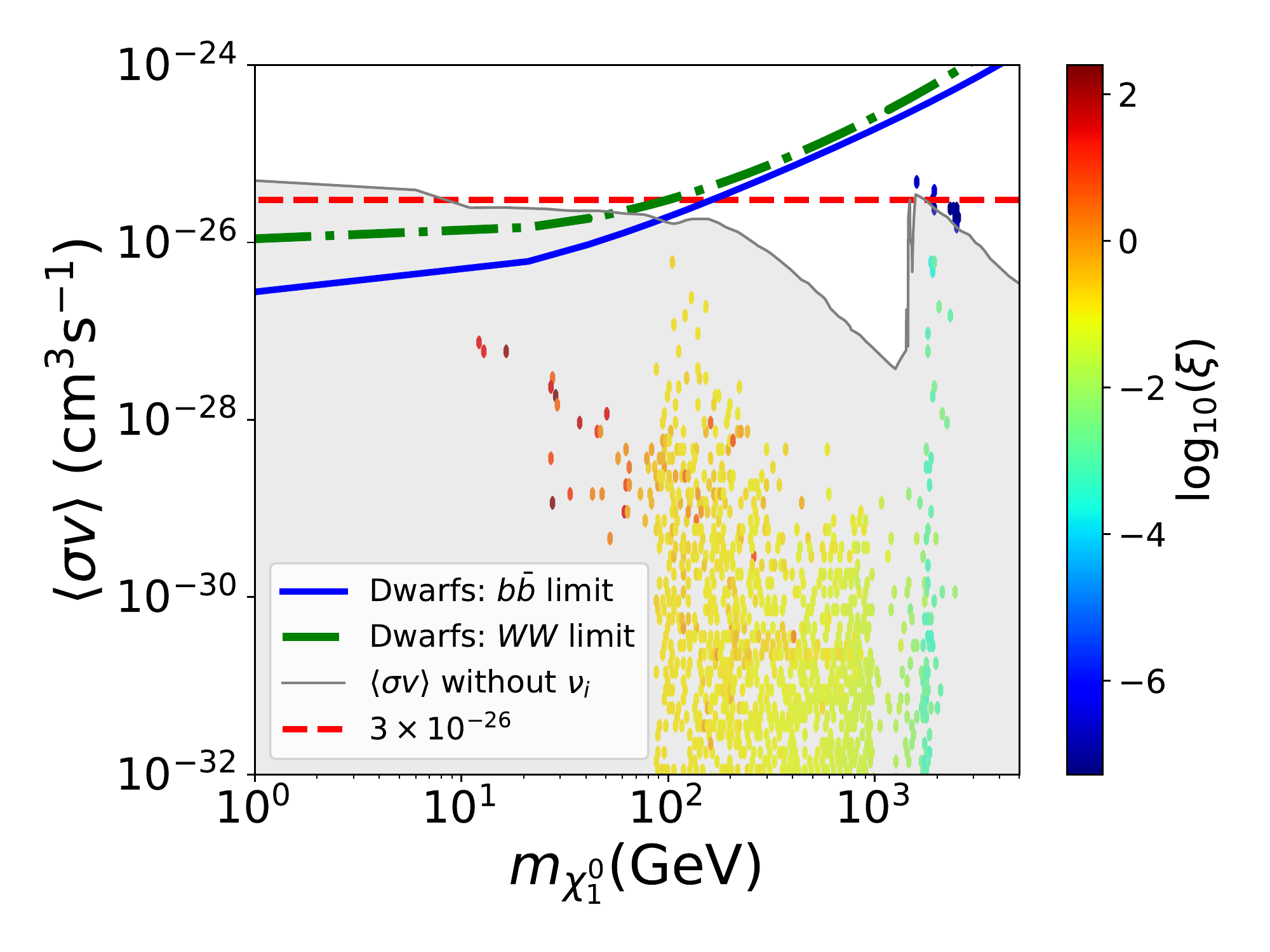}
\caption{Left: SI cross-section. We show the currents limits of XENON1T~\cite{Aprile:2018dbl}, PandaX~\cite{Cui:2017nnn}, and the prospects from LZ~\cite{Akerib:2018lyp} and DARWIN~\cite{Aalbers:2016jon}. We also show the Neutrino Coherent Scattering (NCS)~\cite{Cushman:2013zza, Billard:2013qya} (yellow region). 
Right:  Velocity-averaged annihilation cross-section and current indirect detection limits in $b\bar{b}$ and $WW$ channels~\cite{Ackermann:2015zua}.
In both plots, we also show the region compatible with the relic density but without the correct Yukawa couplings that reproduce the neutrino oscillation parameters (grey region). In colors, we also show the $\xi$ variable defined in Eq.~\eqref{eq:xi}.
}
\label{fig:SI-and-sv-with-neutrinos-scan}
\end{center}
\end{figure}

We computed the SI cross-section ($\sigma_{SI}$) for each point of the scan that was compatible with the relic density of the DM and the neutrino physics. Furthermore, we did a cross-check with the \texttt{MicrOMEGAs 4.2.5} routine~\cite{Belanger:2006is}. 
Our results are shown in the left plot of Fig.~\ref{fig:SI-and-sv-with-neutrinos-scan} together with the current experimental limits of XENON1T~\cite{Aprile:2018dbl}, PandaX~\cite{Cui:2017nnn} and the prospects from LZ~\cite{Akerib:2018lyp} and DARWIN~\cite{Aalbers:2016jon}. 
After this, we clearly see that the scan prefers the region with low $\sigma_{SI}$ which is not currently excluded by the experimental searches of DM. Even more, the majority of the points fall into the Neutrino Coherent Scattering (NCS)~\cite{Cushman:2013zza, Billard:2013qya}, where they will be challenging to looking for in the future~\cite{Mohamadnejad:2019vzg}.
Perhaps, the most important feature is that the neutrino oscillation parameters drastically restring the parameter space of the STFDM model creating a suppression in the $\sigma_{SI}$. 
After the Casas-Ibarra routine described Sec.~\ref{sec:neutrino-masses}, the STFDM model gives us Yukawa couplings $Y_{\Sigma}^i$ and $Y_N^i$ all of them in the range $10^{-5}<|Y_{\Sigma, N}^i|<1$. By construction, they reproduce the neutrino physics and they reduced drastically the parameter space of the first proposal of the STFDM model. In order to show that, we plot in grey the contour of the naked parameter space that is only compatible with DM which was established in Ref.~\cite{Hirsch:2013ola}.   

We also used the \texttt{MicrOMEGAs 4.2.5} routine~\cite{Belanger:2006is} to compute the velocity annihilation cross-section $\langle \sigma v\rangle$ of the  STFDM model for each point of the scan that was compatible with the relic density of the DM and the neutrino physics. 
It is shown in the right side of Fig.~\ref{fig:SI-and-sv-with-neutrinos-scan} with the $95\%$ C.L. gamma-ray upper limits from Dwarf Spheroidal Galaxies (dSphs) for DM annihilation into $b\bar{b}$ and $WW$ channels~\cite{Ackermann:2015zua}.
As in the previous analysis, we also plot the contour of the naked parameter space that is only compatible with DM~\cite{Hirsch:2013ola}.
After this analysis, we realize that the parameter space of the STFDM model is strong reduced when we take into account the neutrino physics.

\subsection{Lepton Flavor Violation}
\label{sec:LFV}

\begin{figure}
\begin{center}
\includegraphics[scale=0.55]{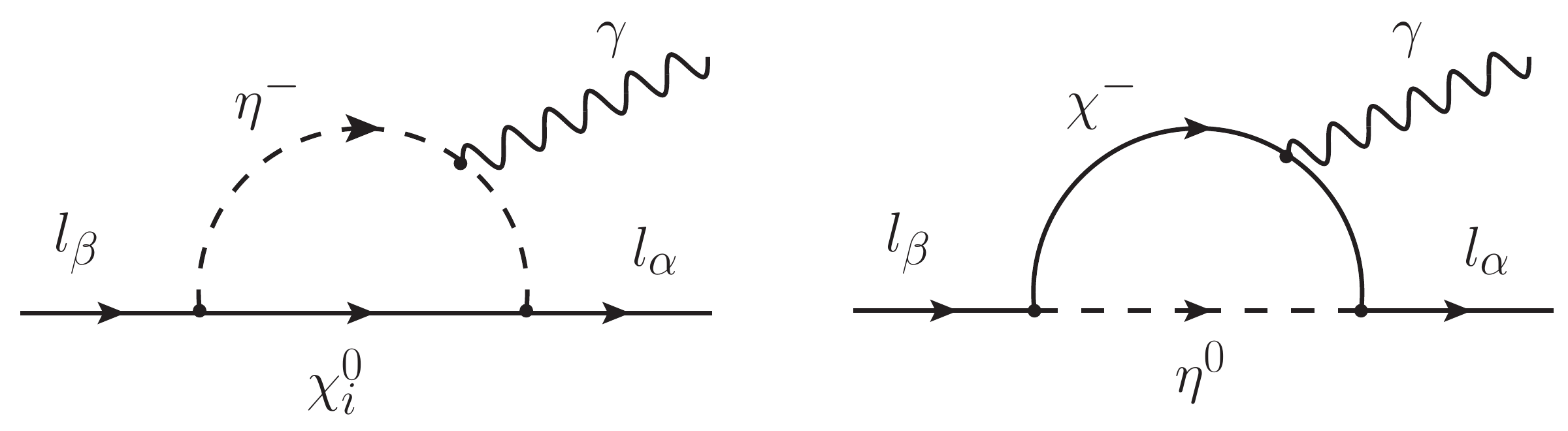}
\caption{Dominant Feynman diagrams in the $l_{\beta}\rightarrow l_{\alpha}\gamma$ process.}
\label{fig:mu-e-gamma}
\end{center}
\end{figure}

The STFDM model allows for lepton flavor violation (LFV) processes that constrain its parameter space. Recently, was shown that the most promising experimental prospects are based on $\mu\rightarrow 3\, e$, $\mu - e$ conversion in nuclei, and 3-body decays $l_{\beta}\rightarrow l_{\alpha}\gamma$, out of which $\mu\rightarrow e\gamma$ is the most important one~\cite{Rocha-Moran:2016enp} (see the Feynman diagrams shown in Fig.~\ref{fig:mu-e-gamma}).

\begin{figure}
\begin{center}
\includegraphics[scale=0.43]{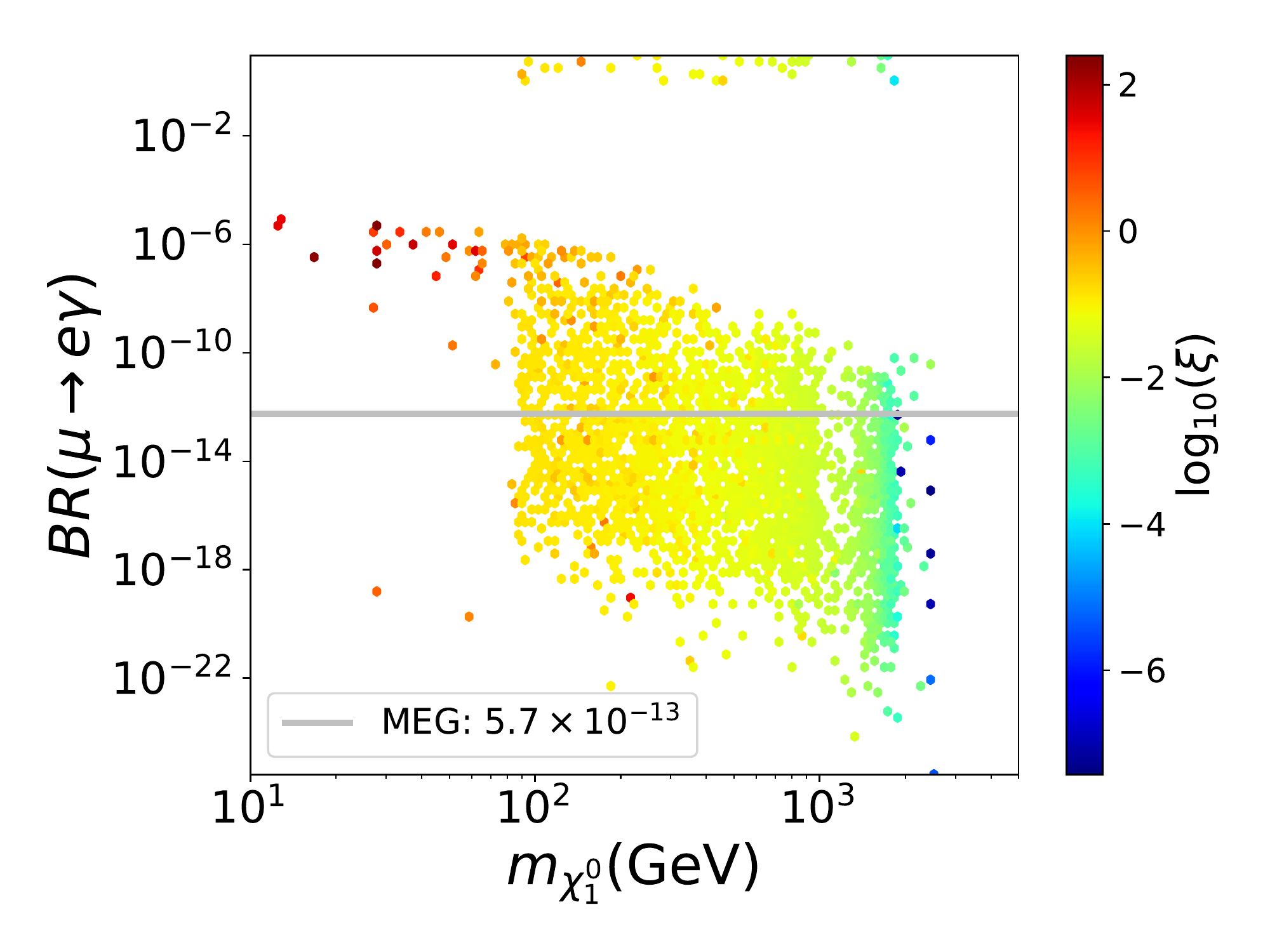}
\includegraphics[scale=0.43]{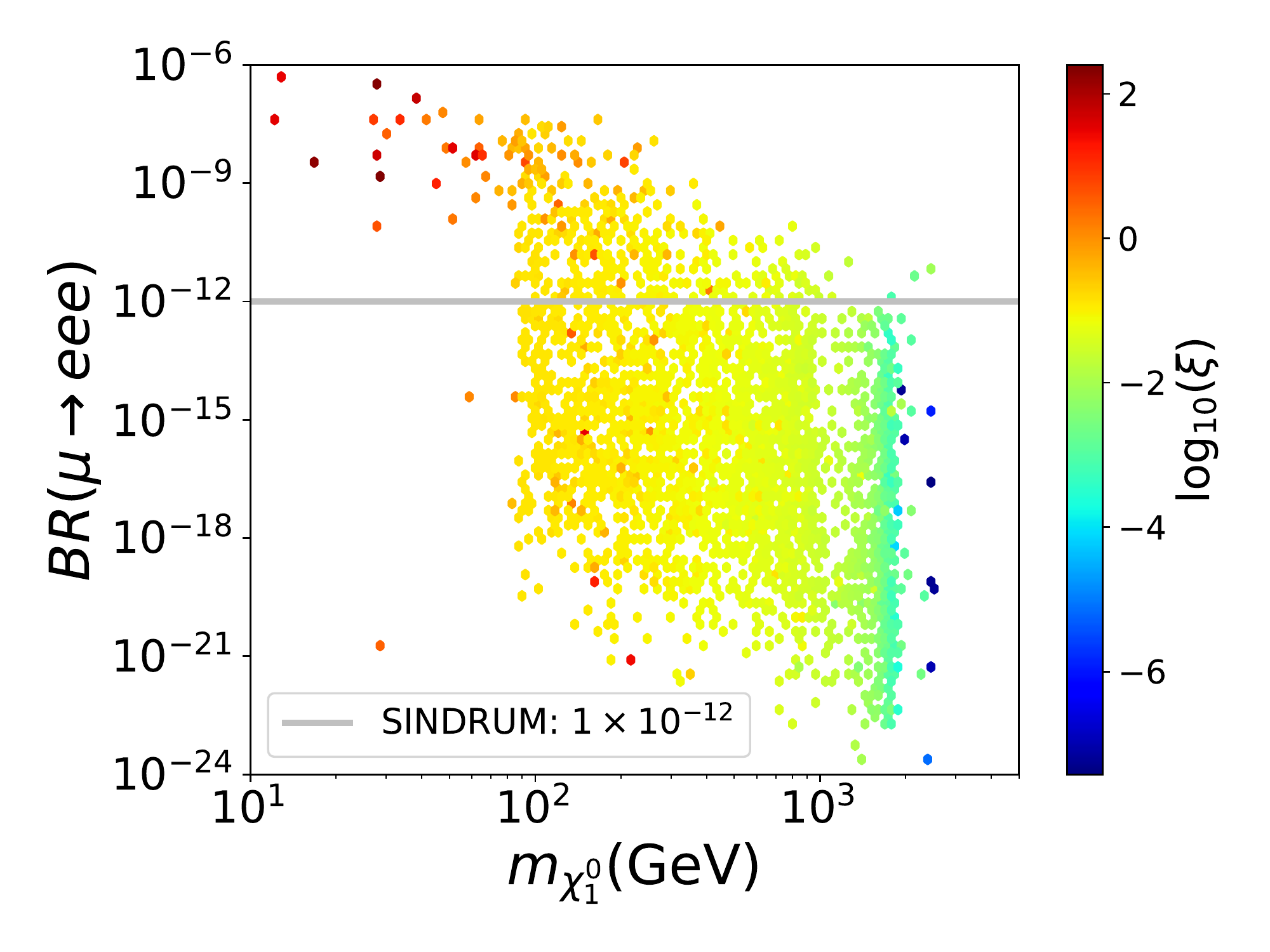}
\caption{Left: $\mu\rightarrow e\gamma$ values for the scan of the parameter space. The grey line show the current limit~\cite{Adam:2013mnn}. The upper region is excluded.
Right: $\mu\rightarrow 3\,e$ values for the scan. The grey line shows the current limit~~\cite{Bertl:1985mw}.}
\label{fig:ueg-e3e}
\end{center}
\end{figure}
On the left side of Fig.~\ref{fig:ueg-e3e}, we show the behavior of the $\mu\rightarrow e\gamma$ process for the scan done in the previous section. 
The analytic expression given in Ref.~\cite{Rocha-Moran:2016enp} was checked with \texttt{FalvorKit}~\cite{Porod:2014xia} of \texttt{SARAH}~\cite{Staub:2008uz,Staub:2009bi,Staub:2010jh,Staub:2012pb,Staub:2013tta} coupled to \texttt{SPheno}~\cite{Porod:2003um,Porod:2011nf} routines. Also, we show the current experimental bounds carried out by the MEG collaboration~\cite{Adam:2013mnn}. 
In addition, we show the $\mu\rightarrow 3\,e$ process and its present bound given by the SINDRUM experiment~\cite{Bertl:1985mw}.
We realize that some points of the parameter space are excluded, especially those with bigger $\xi$ values in the low mass region. We can see that although LFV processes exclude almost all the region with $m_{\chi^0_1}\lesssim 100$~GeV, the majority of the models with $m_{\chi^0_1} \gtrsim 100$~GeV survive and the previous analysis does not change significantly. 
Also, in future, the addition of $\mu - e$ conversion in nuclei process could put new constraints to the STFDM model~\cite{Carey:2008zz,Glenzinski:2010zz,Abrams:2012er,Cui:2009zz,Kuno:2013mha,Barlow:2011zza,Aoki:2010zz,Natori:2014yba}. However, as was shown in Ref.~\cite{Rocha-Moran:2016enp}, that currents bounds of $\mu - e$ conversion in nuclei~\cite{Bertl:2006up} does not put relevant restrictions in this model.

\begin{figure}
\begin{center}
\includegraphics[scale=0.4]{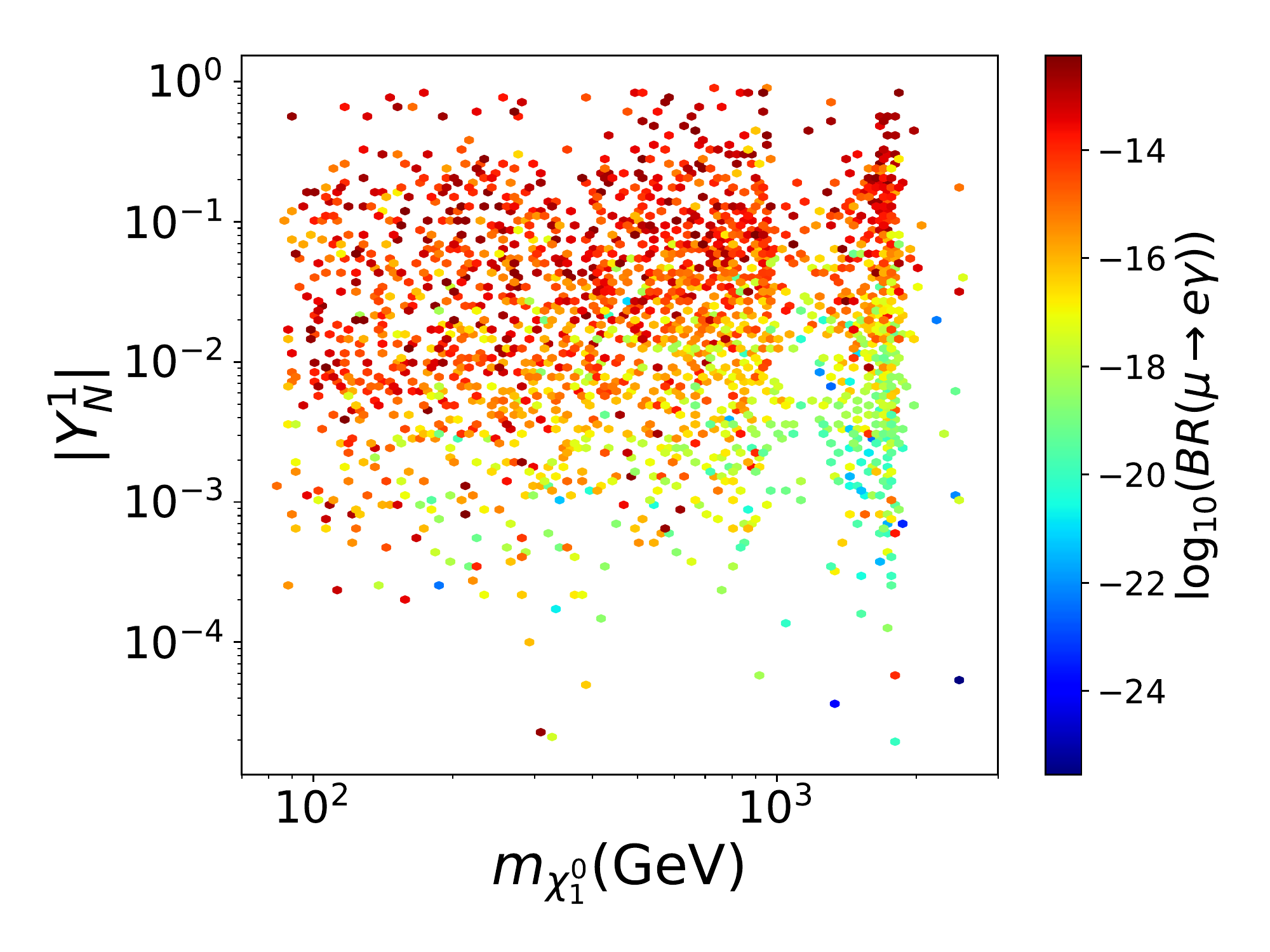}
\includegraphics[scale=0.4]{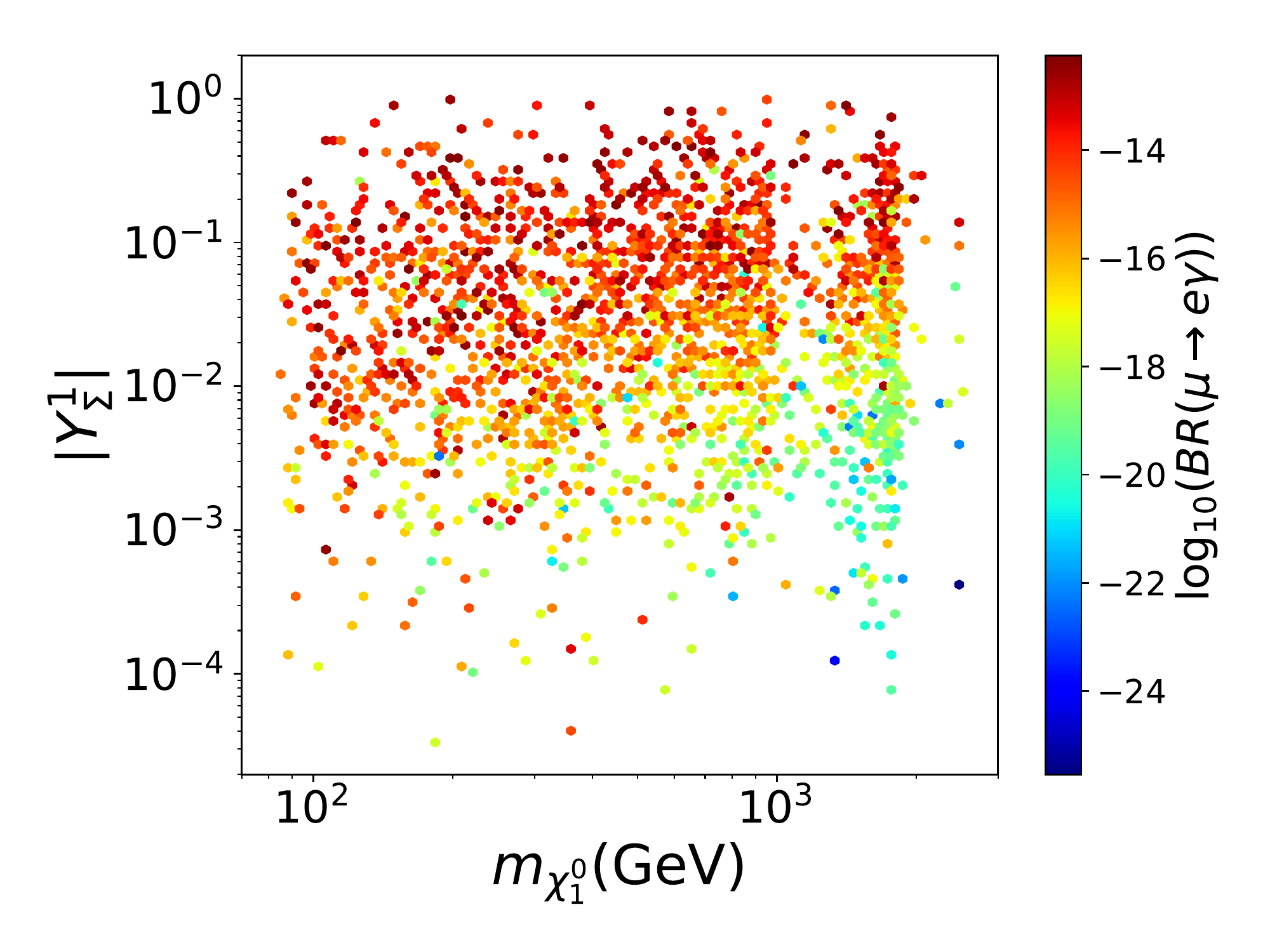}
\includegraphics[scale=0.4]{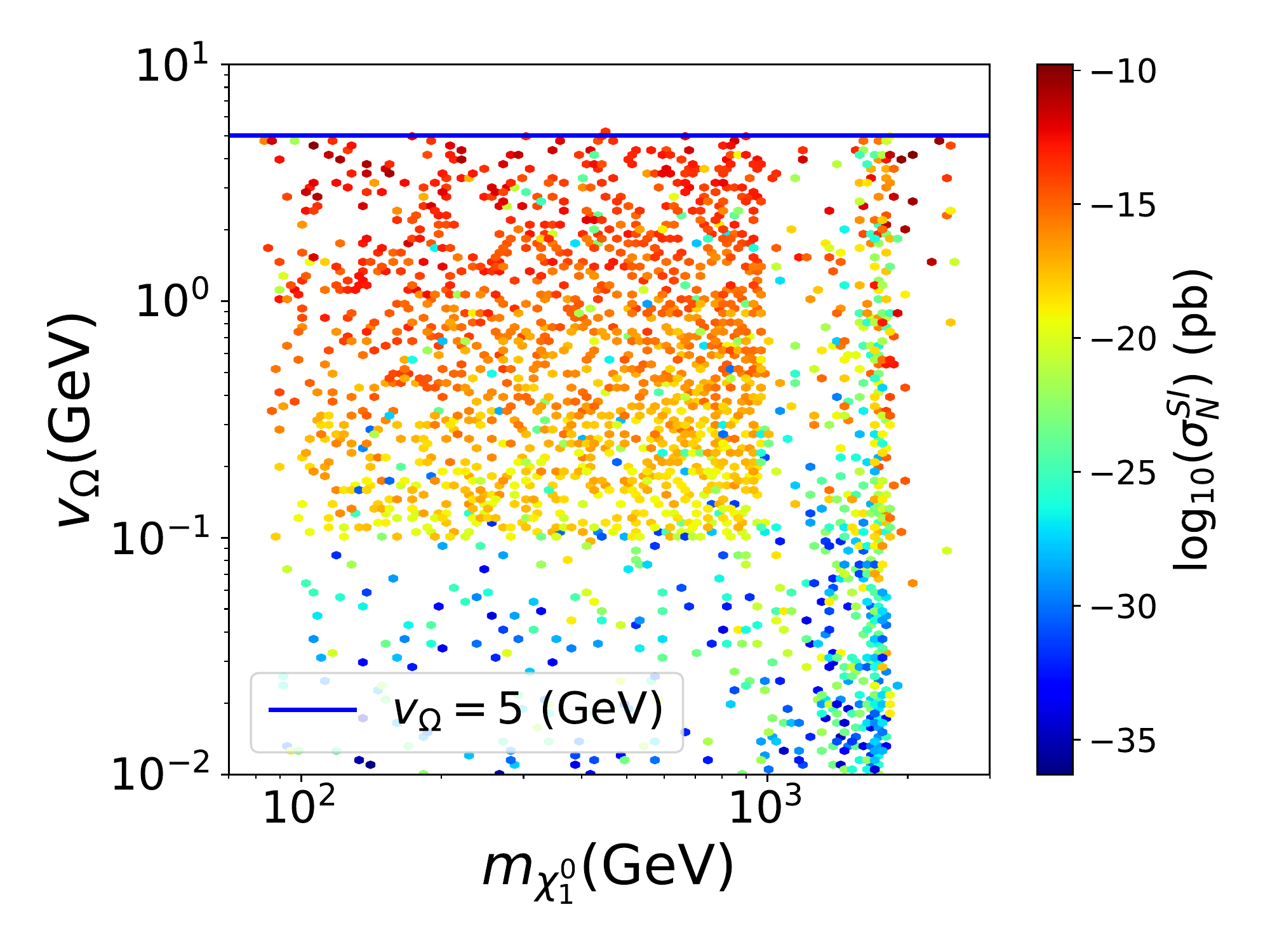}
\includegraphics[scale=0.4]{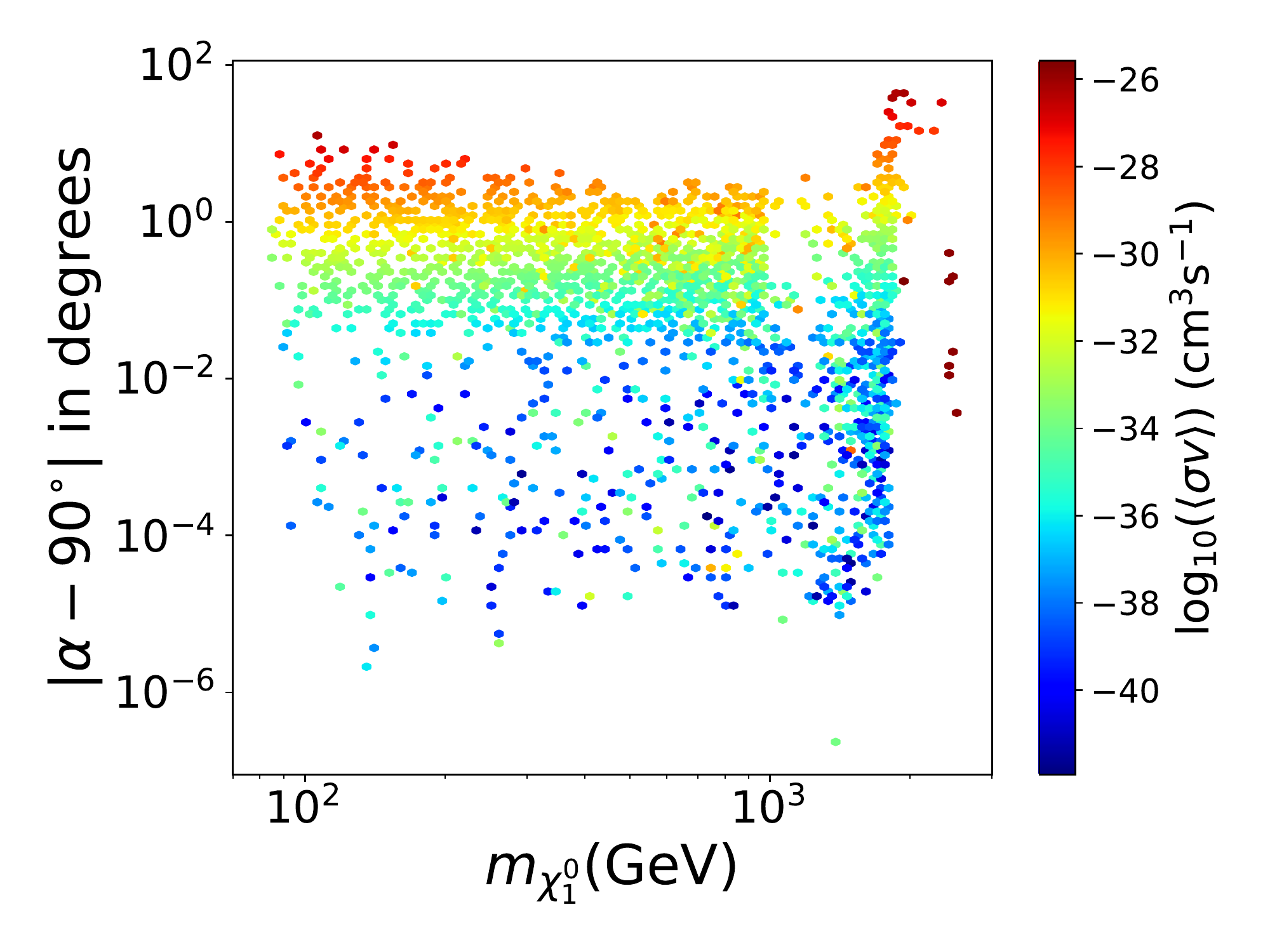}
\caption{Behavior of the parameters of the STFDM model that fulfill DM, neutrino physics and LFV processes.}
\label{fig:model-parameters}
\end{center}
\end{figure}

Finally, in Fig.~\ref{fig:model-parameters} we show the behavior of some of the parameters of the STFDM model that pass all the constraints.
According to those plots, we can draw some conclusions. The Yukawa couplings $Y_{N}^1$ and $Y_{\Sigma}^1$ control the $\mu\rightarrow e\,\gamma$ process. Couplings larger than one give us LFV in the STFDM model. A similar behavior is found for all the Yukawa couplings $Y_{N}^i$ and $Y_{\Sigma}^i$.
The VEV $v_{\Omega}$ of the triplet scalar controls the SI cross-section as we expect by the construction of the STFDM model. 
The velocity-averaged annihilation cross-section is clearly controlled by the mixing angle $\alpha$ defined in Eq.~\eqref{eq:M-chi-rotation} .
Sizable values for $|\alpha-90^{\circ}|$ give us significant values for $\langle\sigma v\rangle$ as we can see in lower-right part of Fig.~\ref{fig:model-parameters}. Those are the promising points of the parameter space that will lead to larger fluxes of gamma-ray as we will show latter.

\subsection{Collider phenomenology}
\label{sec:collider}

\begin{figure}[h]
\begin{center}
\includegraphics[scale=0.6]{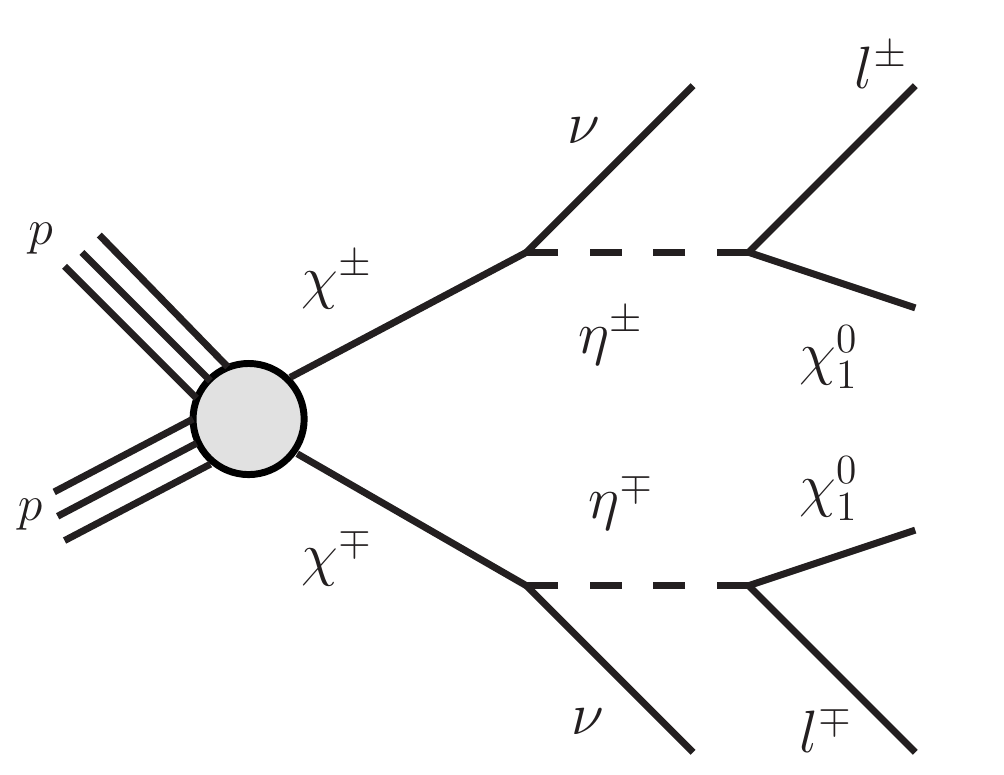}
\includegraphics[scale=0.6]{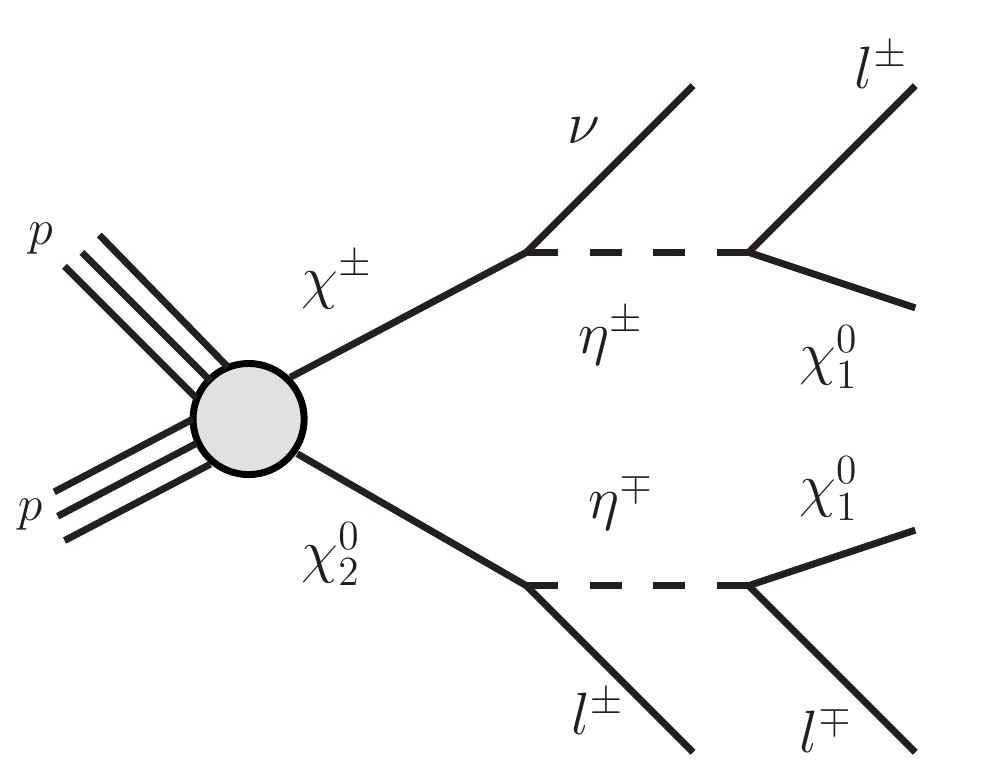}
\caption{Diagrams for two and three leptons plus missing energy that resemble SUSY scenarios. The role of the sleptons is played for the  $\mathbb{Z}_2$-odd scalar $\eta^{\pm}$.}
\label{fig:collider-diagrams}
\end{center}
\end{figure}

We can derive limits on the masses of the new particles of the STFDM model from existing LHC analysis in the context of simplified SUSY models. 
Specifically, we used the ATLAS analysis which constraints the masses for the fermions $\chi^{\pm}$ and  $\chi_i^0$, obtained from searches of wino-like neutralino in the SUSY models~\cite{Aaboud:2018jiw} with decay patterns similar to the those of the STFDM model. Those are shown in Fig.~\ref{fig:collider-diagrams}.
\begin{figure}[h]
\begin{center}
\includegraphics[scale=0.43]{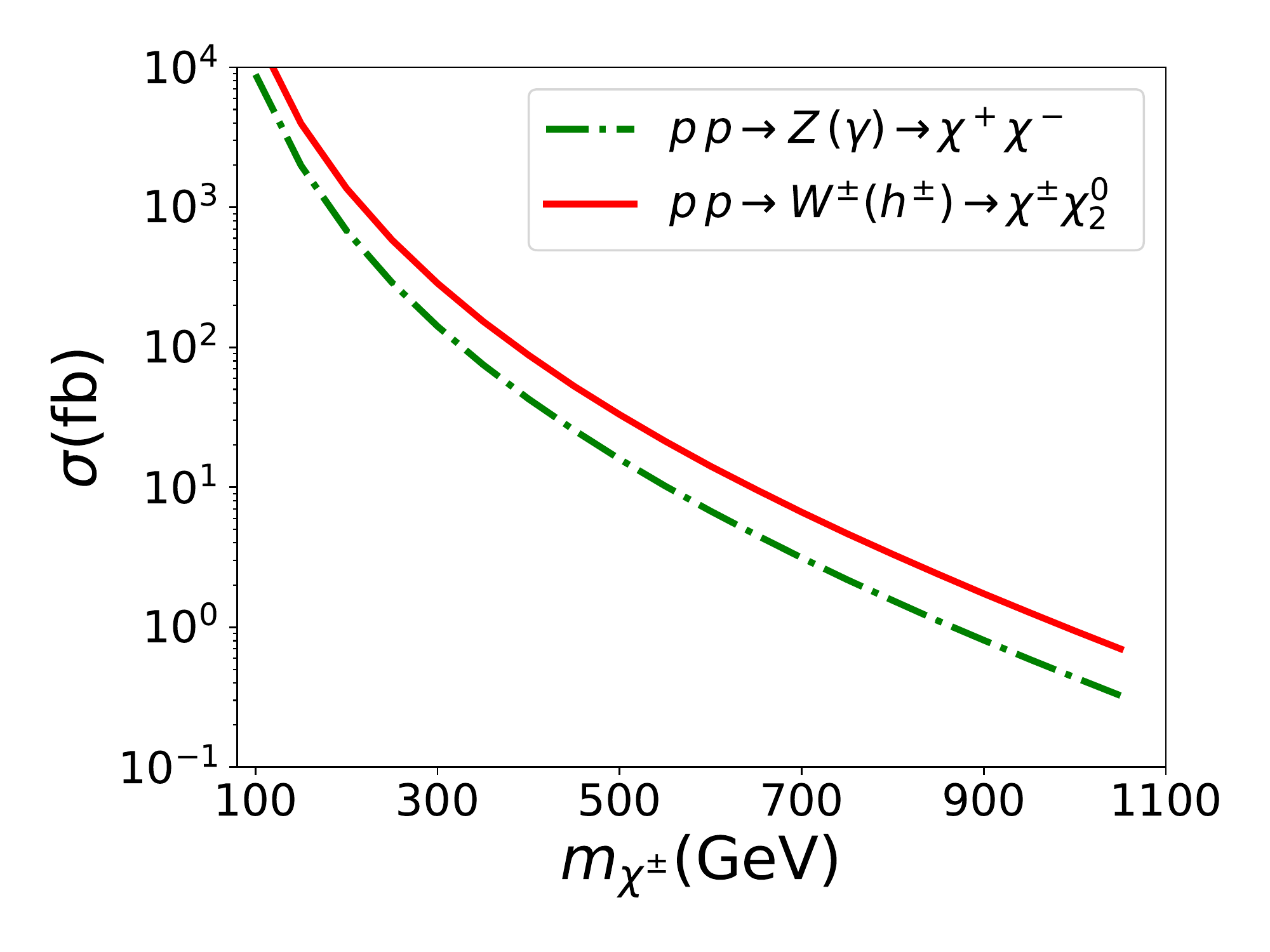}
\includegraphics[scale=0.42]{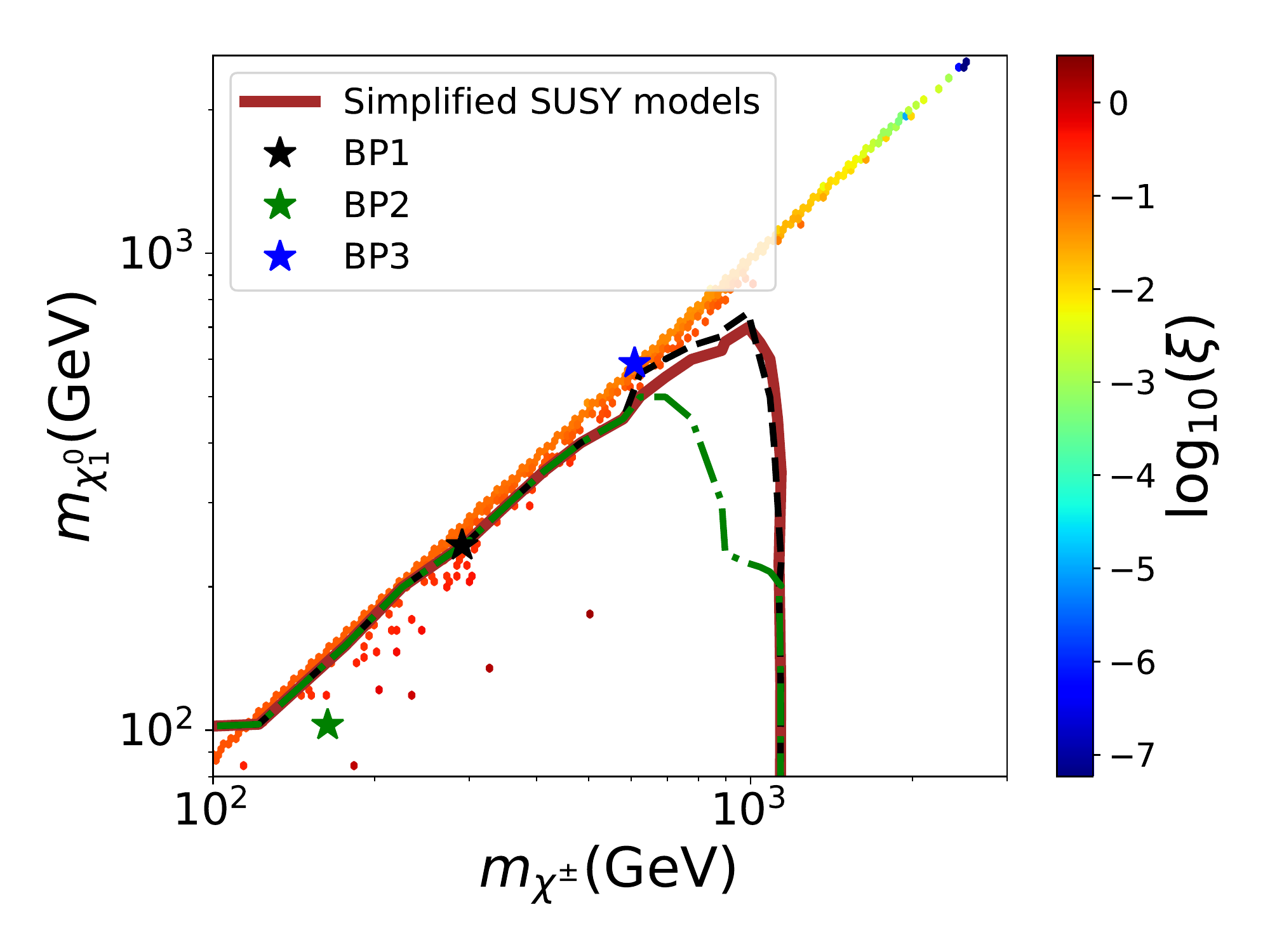}
\caption{ (Left) Production cross-section of $\chi^+\chi^-$, $\chi^{\pm}\chi_2^0\,$ pairs. We fixed $m_{H_2^{\pm}}=300$~GeV and $m_{\chi_2^0}-m_{\chi_1^0}=20$~GeV. 
(Right) Simplified SUSY model, BPs, and the scan discussed in Sec.~\ref{sec:full-scan}. Also, we show the black dashed line and the green dashed-doted line which correspond to the recasting of the ATLAS data for scenarios as BP1 and BP2. The ATLAS limit is based on $36.1\; \text{fb}^{-1}$ of $\sqrt{s} = 13$~TeV proton-proton collisions at the LHC with two or three lepton final state~\cite{Aaboud:2018jiw}.}
\label{fig:LHC}
\end{center}
\end{figure}
In general, the DM production is associated to the production cross-section of the processes $p\,p\to\chi^+\chi^-$, $p\,p\to\chi^{\pm}\chi_2^0\,$ and $p\,p\to\chi_2^0\chi_2^0\,$. 
In Fig.~\ref{fig:LHC} (left) we show the production cross-section of the first and the second processes.
Those were computed with \texttt{MadGraph5(v2.5.5)}\cite{Alwall:2014hca} to leading order.
The $\chi_2^0\chi_2^0$ pair production is not showed because is very small compared to the other two processes. 
We see that the production cross-section of $\chi^{\pm}\chi^0_2$ pair is bigger compared to the $\chi^{+}\chi^{-}$ pair.
However, the $\chi^{\pm}\chi^{\mp}$ pair production is a cleaner channel at the LHC. In the first case, the $\chi_2^0$ fermion dominantly decay into $b\bar{b}$ pair along with $\chi_1^0$ as was argued in Ref.~\cite{Choubey:2017yyn}. We choose these channels to do our analysis.

The idea in this section is to show that, although some points of this model resemble the SUSY scenario, they need to be recast because they do not fulfill completely the assumptions of the simplified SUSY models.
First of all, we have to take into account that in SUSY simplified models it is assumed that the chargino ($\tilde{\chi}_1^{\pm}$) decays into neutrinos and sleptones ($\tilde{l}$) with a branching ratio BR$(\tilde{\chi}_1^{\pm}\to \nu\,\tilde{l})=50\,\%$. The other half decays directly into leptons and sneutrinos $(\tilde{\nu})$.
At the same time, is assumed that the sleptons decay completely to electrons and muons together with the lightest neutralino with a BR$(\tilde{l}\to l\,\tilde{\chi}_1^0)=100\,\%$.
However, in the STFDM model, is difficult to satisfy those assumptions because the vertices of those processes are given by the Yukawa couplings $Y_N^{\alpha}$ and $Y_{\Sigma}^{\alpha}$ of the Lagrangian~\eqref{eq:full-lagrangian} that are controlled by the restrictions imposed by the Casas-Ibarra parametrization of neutrino physics~\cite{Casas:2001sr, Ibarra:2003up}.
Taken this into account, in Table~\ref{tab:B-Points} we show some benchmark points (BP) of this model.
First, the BP1 partially fulfill the SUSY assumptions where the scalar $\eta^{\pm}$ decay almost completely to electrons together with the lightest fermion of the STFDM model (the DM particle) with a branching ratio BR$(\eta^{\pm}\to e^{\pm}\,\chi_1^0)\sim 98\,\%$. 
However, the $\text{BR}(\chi^{\pm}\to \nu\, \eta^{\pm})\sim 100\,\%$ and therefore the cross-section given by ATLAS needs to be rescaled by a factor 2 for each vertex with the neutrino.
Secondly, we show the BP2, where the final leptons states are not $100\,\%$ muons or electrons. It escapes partially the SUSY analysis because there is a $\sim50\,\%$ of tau leptons and therefore the cross-section given by the ATLAS analysis needs to be rescaled by a factor $1/2$ for each vertex with charged leptons.
As a final benchmark point, we show the BP3 which escapes completely the SUSY analysis. In this case, the final state are mainly tau leptons with a BR$(\eta^{\pm}\to \tau^{\pm}\,\chi_1^0)\sim 89\,\%$, which is not considered in ATLAS analysis.
\begin{table}
\centering
\resizebox{\textwidth}{!}{
\begin{tabular}{|c|c|c|c|c|c|c|c|}
\hline
 &&&&&&& \\   
 & $m_{\chi^{\pm}}[\text{GeV}]$ & $m_{\chi_2^0}[\text{GeV}]$ & $m_{\eta^{\pm}}[\text{GeV}]$ & $m_{\chi_1^0}[\text{GeV}]$ & $\text{BR}(\chi_2^0\to \eta^{\pm}l^{\mp})$& $\text{BR}(\chi^{\pm}\to \nu\, \eta^{\pm})$ & $\text{BR}(\eta^{\pm}\to l^{\pm}\, \chi_1^0)$  \\
\hline
 &&&&&&& \\   
BP1 & 290.7 & 290.7  & 255.2 & 242.2 & $ 50\,\%$ to $e^{\pm}$ & $ 100\,\%$  & $ 98\,\%$ to $e^{\pm}$\\
    &  &  &  &  & $ 50\,\%$ to $\mu^{\pm}$ &  &$ 2\,\%$ to $\mu^{\pm}$ \\   
 &&&&&&& \\   
BP2 & 163.4 & 163.4  & 107.6 & 102.6 & $50\,\%$  to $e^{\pm}$  & $ 50\,\%$ & $ 50\,\%$ to $\mu^{\pm}$\\
     & &  &  &  &    &  & $ 50\,\%$ to $\tau^{\pm}$\\    
 &&&&&&& \\        
BP3 & 608.5 & 608.5  & 598.3 & 588.1 & $ 35\,\%$ to $\mu^{\pm}$  & $ 50\,\%$ & $ 8\,\%$ to $\mu^{\pm}$\\  
 &  &   &  &  & $ 15\,\%$ to $\tau^{\pm}$  &  & $ 89\,\%$ to $\tau^{\pm}$\\ 
   \hline 
\end{tabular}}
\caption{Benchmark points to look at for collider signals.}
\label{tab:B-Points}
\end{table}

In Fig.~\ref{fig:LHC} (right), we show the LHC analysis in the context of simplified SUSY models (brown line). Those are projected on the plane of $m_{\chi^{\pm}}$- $m_{\chi_1^0}$ as usually done in ATLAS plots. We also show the three BPs and the scan done in Sec.~\ref{sec:full-scan}.
To complement this analysis, we also show the recasting of the ATLAS data for models as BP1 and BP2 (black dashed and green dashed-doted line). In this procedure, we rescaled the ATLAS cross-section appropriately as we described before.
In the end, we find that collider searches could test masses up to $\sim 700$ GeV in the most conservative cases. However, it is challenging because we have compressed spectra and a better analysis needs to be done in this direction and we leave it for future work.

\section{One-loop prospective observables}
\label{sec:1-loop-processes}

In this section, we compute some new observables that arise at one-loop level in the STFDM model. These are the SD cross-section of DM recoil with nuclei and the DM annihilation into two photons. Both of them are promising process for future signals of this model.

\subsection{Spin-dependent cross-section at one-loop}
\label{sec:sigma-SD}

\begin{figure}[h]
\begin{center}
\includegraphics[scale=0.45]{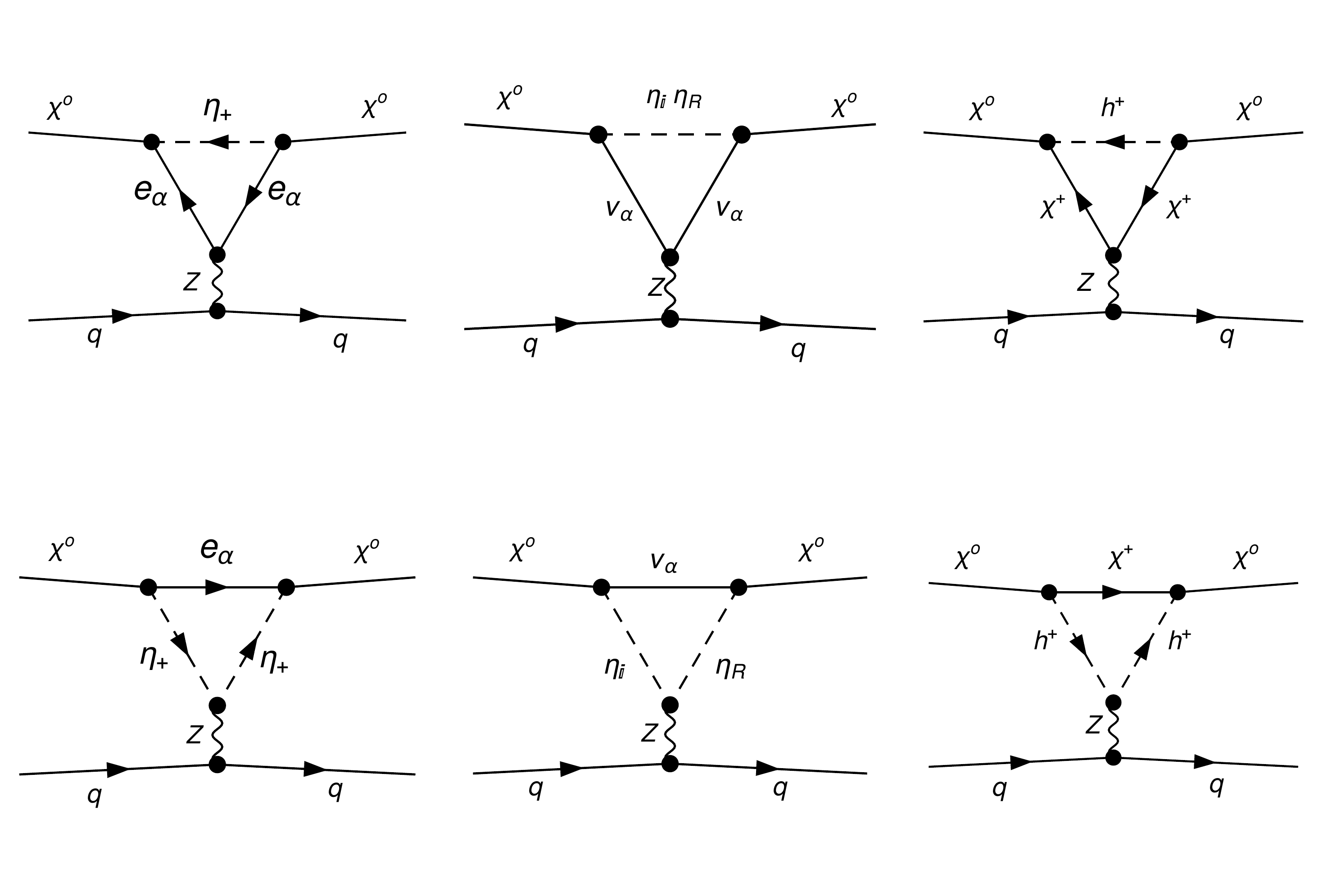}
\caption{Diagrams contributing to SD cross-section at one-loop level. They were generated using \texttt{FeynArts}~\cite{Hahn:2000kx}.}
\label{fig:SD}
\end{center}
\end{figure}
Although the STFDM model is blind to SD scattering of DM at tree-level, the scattering can occur at one-loop level as shown in Fig.~\ref{fig:SD} (we only show diagrams with charged particles circulating in one direction). 
Concretely,  the exchange of the  $Z$ boson leads to an effective axial vector interaction term of the form~\cite{Jungman:1995df, Ibarra:2016dlb}
\begin{align}
\label{eq:LSD-eff}
\mathcal{L}_{\text{eff}}\, =\, \xi_q \bar{\chi^0_1}\gamma^{\mu}\gamma^5 \chi^0_1 \bar{q}\gamma_{\mu}\gamma^5q + \text{h.c.}\,,
\end{align}
where
\begin{align}
\label{eq:xiq}
\xi_q = \dfrac{a_q \sin^2\alpha }{32\pi^2M_Z^2}
\left\{
\sum_{\alpha}|Y_N^{\alpha} |^2\left[
(v_{e}+a_{e}) \mathcal{G}_2\left(\dfrac{m_{\chi^0_1}^2}{m_{\eta^{\pm}}^2}\right) 
+ (v_{\nu}+a_{\nu}) \mathcal{G}_2\left(\dfrac{m_{\chi^0_1}^2}{m_0^2}\right) 
\right]
+ 
Y_{\Omega}^2(v_{\chi}+a_{\chi}) \mathcal{G}_2\left(\dfrac{m_{\chi^0_1}^2}{m_{h_2^+}^2}\right) 
\right\}\,,
\end{align}
with $a_q = \dfrac{1}{2}\,$ for $q=(u,c,t)$,
$a_q = -\dfrac{1}{2}$ for $q=(d,s,b)$, 
$a_{e} = -\dfrac{g}{2c_W}\left(\dfrac{1}{2}\right)$, $v_{e} = -\dfrac{g}{2c_W}\left(\dfrac{1}{2}-2s^2_W \right)$, $a_\nu = v_\nu = \dfrac{g}{2c_W}\left(\dfrac{1}{2}\right)$, $a_{\chi} = 0$, $v_{\chi} = -g\, c_W$, and $m_0\approx m_{\eta^R}\approx m_{\eta^I}$. $\mathcal{G}_2(z)$ is a loop function given by
\begin{align}
\label{eq:G22}
\mathcal{G}_2(z) = -1 + \dfrac{2(z+(1-z)\ln(1-z))}{z^2}\,.
\end{align}
The resulting SD cross-section per nucleon  $N$  is given by
\begin{equation}
\label{eq:SD}
\sigma_N^{SD}=\dfrac{16}{\pi}\dfrac{m_{\chi^0_1}^2m_N^2}{(m_{\chi^0_1}+m_N)^2}J_N(J_N+1)\left(\sum_{q=u,d,s}\Delta_q^N \xi_q\right)^2 \,,
\end{equation}
where $\Delta_u^N\approx 0.842$, $\Delta_d^N\approx -0.427$ and $\Delta_s^N\approx -0.085$~\cite{Airapetian:2006vy}, and $m_N$ and $J_N$ are the mass and angular momentum of the nucleus. 
Notice that we have two contributions to the $\xi_q$ effective coupling. 
The first one is proportional to $Y_N^{\alpha}$ and is common to the original scotogenic model~\cite{Ma:2006km}. The second one, with the charged fermion $\chi^{+}$ and proportional to $Y_{\Omega}^\alpha$, is characteristic of the STFDM model and could enhance the SD cross-section. We checked that, in the limit of $\alpha\sim\pi/2$ and $Y_{\Omega}^\alpha=0$, we recovered the results found in Ref.~\cite{Ibarra:2016dlb}.

\begin{figure}[h]
\begin{center}
\includegraphics[scale=0.5]{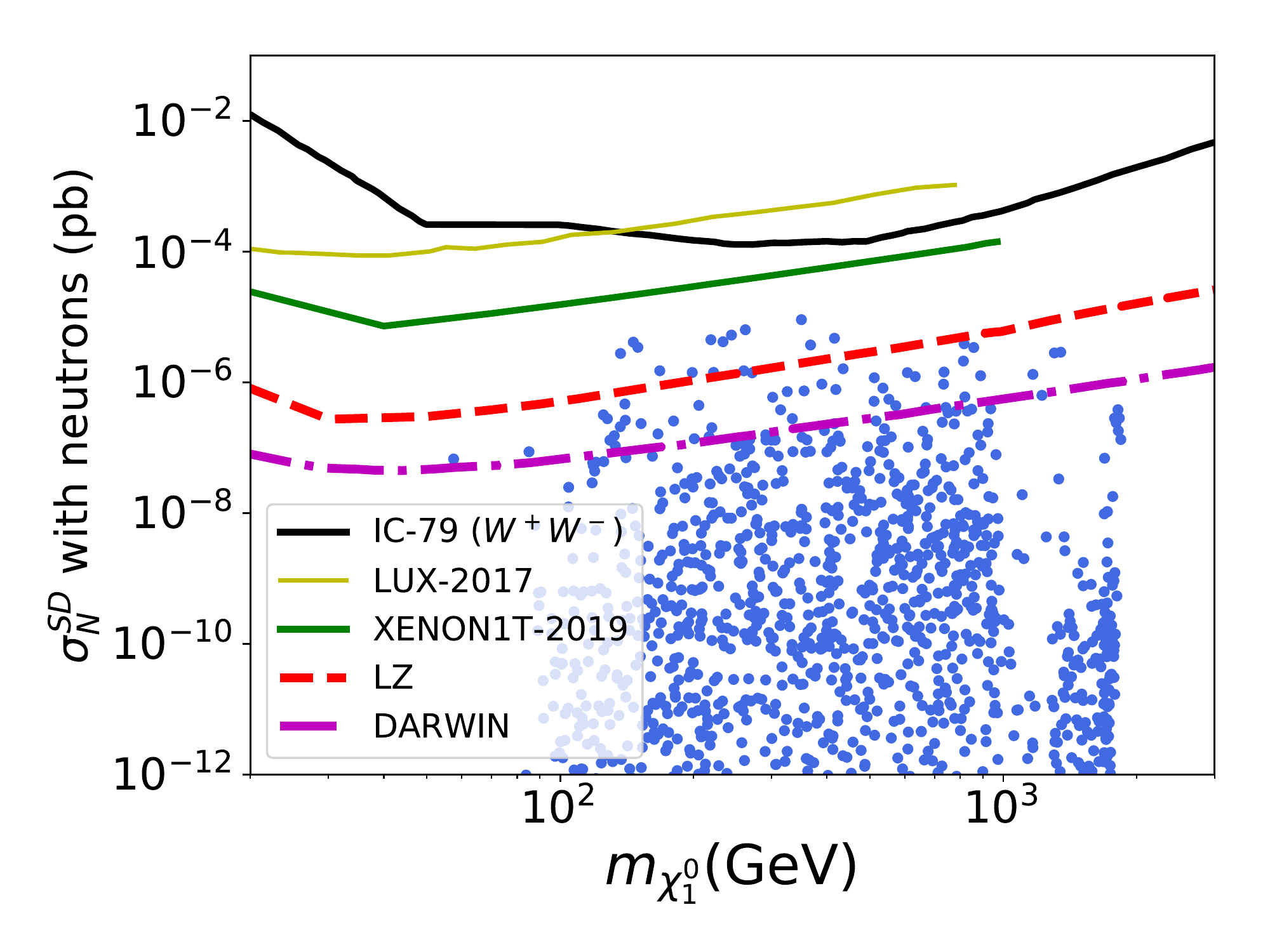}
\caption{SD cross-section with its current limits and future prospects. }
\label{fig:SD-scan}
\end{center}
\end{figure}

In Fig.~\ref{fig:SD-scan}, we show the behavior of the WIMP-neutron SD cross-section for all the models found in the previous section that yield the expected value of the relic abundance, the correct neutrino oscillation parameters, and are not excluded by LFV processes.
We also show the IceCube~\cite{2013PhRvL.110m1302A} limits in the $W^+W^-$ channel (black solid line) for DM annihilation at the sun, the limits from LUX~\cite{Akerib:2016lao} (yellow solid line), the current limits from XENON1T~\cite{Aprile:2019dbj} (green solid line) and the expected sensitivity of LZ~\cite{Akerib:2018lyp} (red dashed line) and DARWIN~\cite{Aalbers:2016jon}(magenta dot-dashed line).
We found that the STFDM model is not excluded by SD scattering of DM with nuclei even by the next generation of experiments, such as LZ and DARWIN.

\subsection{Gamma-ray signal: DM annihilation into two photons}
\label{sec:gamma-ray}
  
In general, the DM annihilation into photons is a loop process involving multiple Feynman diagrams. 
It is an interesting process because it could produce a mono-energetic spectral line that would be a strong indication of the existence of the DM. We know that this line-like spectrum is quite difficult to explain using the known astrophysical objects in the universe, and for that reason, its finding would be a clear hint of DM (For a review, see Ref.~\cite{Bertone:2004pz}).
     
In the STFDM model, the DM could annihilate into two photons ($\chi_1^0\chi_1^0\rightarrow\gamma\gamma$) and into photon plus $Z$ gauge boson ($\chi_1^0\chi_1^0\rightarrow\gamma Z$). 
However, in this work, we only computed the amplitude for the first process, the latter one is out of the scope of this work.
Following the general expression given in Ref.~\cite{Garcia-Cely:2016hsk}, we computed the general amplitude for the $\chi_1^0\chi_1^0\rightarrow\gamma\gamma$ process. 
Also, we used \texttt{FeynArts}~\cite{Hahn:2000kx} and \texttt{FormCalc} to reduce the tensor loop integrals to scalar Passarino-Veltman functions~\cite{Passarino:1978jh} and we used \texttt{Package-X}~\cite{Patel:2015tea} to compute the amplitude of this process~\footnote{In the Feynman gauge there are 340 Feynman diagrams for one lepton family of the SM which are classified according to the topologies described in Ref.~\cite{Garcia-Cely:2016hsk}.}. 
Finally, we did a cross-check between these two techniques.  

The cross-section for this process is given by
\begin{align}
\label{eq:sigmav-gg}
\sigma v (\chi^0_1\chi^0_1\rightarrow\gamma\gamma) &= \dfrac{|\mathcal{B}|^2}{32\pi m_{\chi^0_1}^2}\,,
\end{align}
where the $\mathcal{B}$ factor is a scalar function that is given in the Appendix~\ref{app:B-factor}, Eq.~\eqref{eq:B}. It was written in such a way that we factorized the gauge invariant contribution in order to see the impact of the different parameters of the STFDM model.
Even more, in the Appendix~\ref{app:B-factor} we show that this general expression reproduces some known limits. 
For instance, in the limit of singlet fermion DM, which is, $\alpha =\pi/2$ and $Y_{\Omega}=0$, the Eq.~\eqref{eq:sigmav-gg} reproduces the amplitude of the original scotogenic model~\cite{Garny:2015wea}. This is shown in Sec.~\ref{sec:pure-singlet-sigmagg}.
In the same way, in the limit of pure triplet DM, that is $Y_{\Omega}=Y_{N}^{\alpha}=Y_{\Sigma}^{\alpha}=v_{\Omega}=0$, $\alpha =0^{\circ}$ and $m=M_{\Sigma}$,  it also reproduces the results obtained in the high mass region for minimal DM model~\cite{Cirelli:2005uq}. This is shown in Sec.~\ref{sec:pure-triplet-sigmagg}.

\begin{figure}
\begin{center}
\includegraphics[scale=0.5]{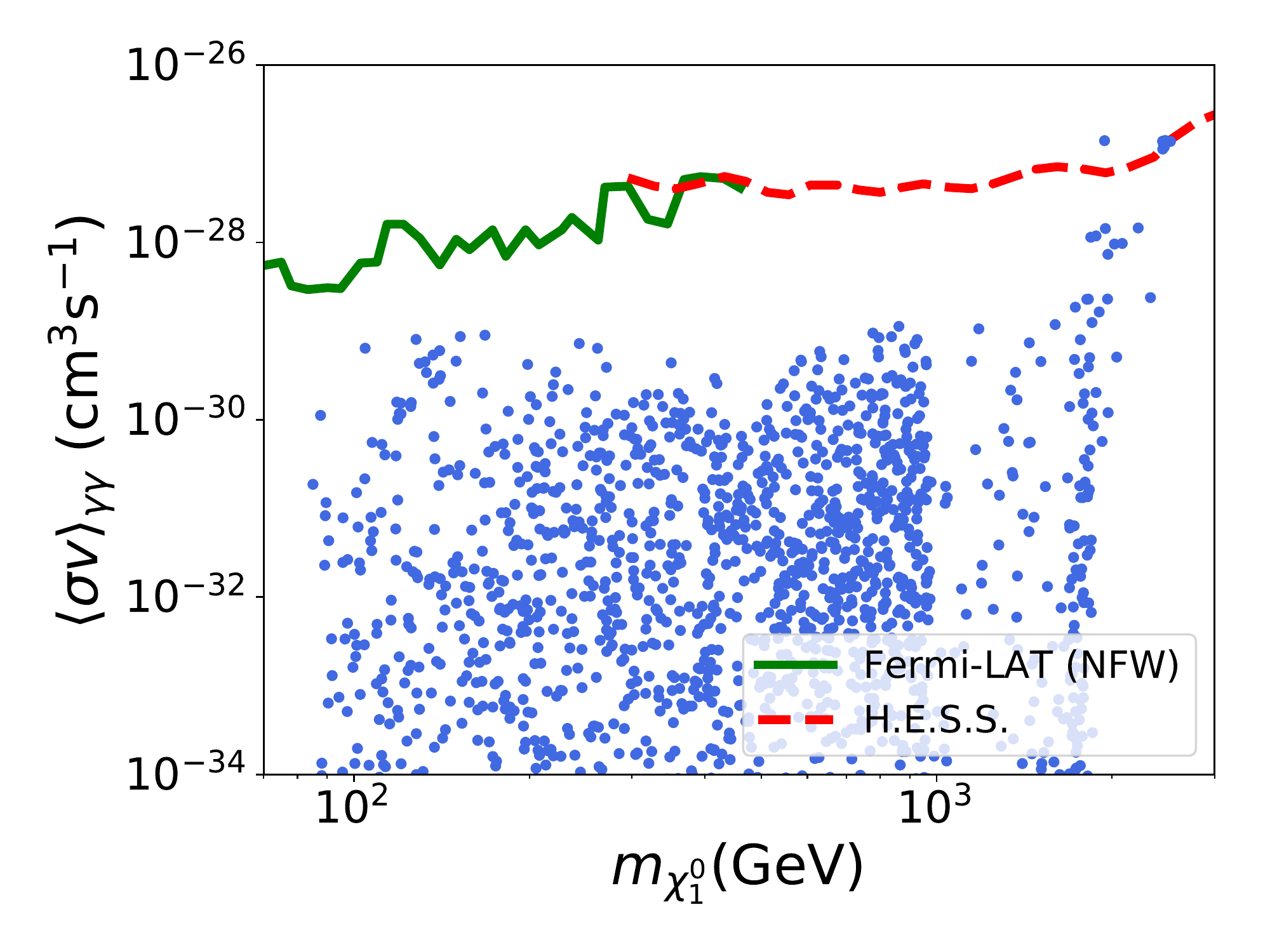}
\caption{DM annihilation into two photons in the STFDM model. The continues (dashed) line represent the current bound of Fermi-LAT (H.E.S.S.) collaboration for DM annihilation into two photons in the Milky Way Galaxy.}
\label{fig:sigmavgg}
\end{center}
\end{figure}

In Fig.~\ref{fig:sigmavgg} we show the DM annihilation into two photons for the scan done in Sec.~\ref{sec:full-scan}. We only show the points which are in agreement with the LFV processes described in Sec.~\ref{sec:LFV}, neutrino physics and yield the expected value of the relic abundance of DM.
We also show the current bounds of the Fermi-LAT~\cite{Ackermann:2015lka} collaboration for observation of the Milky Way halo in the low mass region $\sim (200\,\text{MeV}-500\,\text{GeV})$ and the H.E.S.S.~\cite{Abdallah:2018qtu} bounds for the high mass region $\sim (300\,\text{GeV}-70\,\text{TeV})$.
After improving our scan as much as possible, we realize that all the points always fall under the Fermi-LAT bound in the low mass region. 
For high masses, the STFDM model reaches the current bound of H.E.S.S., however, those points were computed for illustration because we were interested in the low mass region. For the limit of masses at the TeV scale in the triplet case, see Ref.~\cite{Cirelli:2007xd}.

\section{Conclusions}
\label{sec:conclusions}

In this paper, we studied  the full consistency of the STFDM model by performing a comparative analysis of a variety of observables. 
We focused on the phenomenology when the DM is the lightest particle that emerges from the mixing between the singlet and triplet fermion.
We studied the parameter space that is fully consistent with the  DM relic abundance while yielding measured parameters of neutrino physics.
In order to achieve this, we randomly scanned the parameter space of the STFDM model imposing a variety of theoretical constraints. 

We realized, although the mixture between the triplet and the singlet fermion is important, the parameters space that is fully consistent with the DM abundance and the neutrino physics, prefers a singlet component $\sim N$ in the low mass region.
Also, we found that coannihilation process between the singlet and the triplet fermion plays an important role and brings the relic density to its observed value for almost all the points with $80~$GeV $<m_{\chi_1^0}<2.4$~TeV. 
In general, we realized that the neutral fermion spectrum is almost degenerate for the majority of the points up to $2.4$~TeV. For masses larger than this value, the STFDM model recovers the known limit of the Minimal DM scenarios in which the DM particle is the triplet fermion. 
We also found that the direct and indirect signals of the model are seriously restricted by neutrino physics constraints. 

Additionally, we complemented the analysis with some LFV processes, such as $\mu\to\,e\,\gamma$ and $\mu\to 3\,e$, and with some searches of DM at the LHC. 
We encountered that DM with a mass in range $80$~GeV $\lesssim m_{\chi_1^0}\lesssim$ $2.4$~TeV is fully consistent and could be tested in future searches of DM. Lighter masses are excluded by LFV processes.

Finally, we computed the SD cross-section of DM at one-loop level and the DM annihilation into two photons ($\chi_1^0\chi_1^0\to\,\gamma\gamma$). As far we know, those two expressions are reported for the first time for this model.
We showed that SD cross-section reaches the future prospects for searches of DM. Specifically, the next generation of experiments as LZ and DARWIN will improve the current limit of XENON1T by up two orders of magnitude and will test a small region of the parameter space for $ m_{\chi_1^0}\lesssim$ $1$~TeV.
On the other hand, we found that DM annihilation into two photons does not further constrain the model. Specifically, the cross-section is $\langle \sigma v\rangle_{\gamma\gamma}\lesssim 10^{-29}$ cm$^3$s$^{-1}$ for $ m_{\chi_1^0}\lesssim$ $1$~TeV, which is below the current limits reported by the Fermi-LAT and H.E.S.S. collaborations.

\section{Acknowledgments} 
We are grateful to Walter Tangarife for reading the manuscript and Oscar Zapata for enlightening discussions. DR is partially  supported by COLCIENCIAS grant 111577657253 and Sostenibilidad-UdeA.
AR is supported by COLCIENCIAS through the ESTANCIAS POSTDOCTORALES program 2017. 

\appendix

\section{General $\mathcal{B}$ factor in the STFDM model}
\label{app:B-factor}

We factorized the $\mathcal{B}$ factor in gauge invariant terms in order to get clarity in the expression
\begin{align}
\label{eq:B}
\mathcal{B} &=
\frac{ \sqrt{2} \alpha_{em}  m^2 \sin ^2(\alpha )Y_{\Omega }^2 (\sin (\delta )+\cos (\delta ))^2}{\pi }
\bigg[
\frac{M_{H^{\pm }}^2 C_0\left(0,-m^2,m^2;M_{H^{\pm }}^2,M_{H^{\pm }}^2,M_{\Sigma }^2\right)}{M_{H^{\pm }}^2-M_{\Sigma }^2}\nonumber\\
&-\frac{M_{\Sigma } \left(-2 m M_{H^{\pm }}^2-M_{\Sigma } M_{H^{\pm }}^2+m^2 M_{\Sigma }+2 m
   M_{\Sigma }^2+M_{\Sigma }^3\right) C_0\left(0,-m^2,m^2;M_{\Sigma }^2,M_{\Sigma }^2,M_{H^{\pm }}^2\right)}{\left(M_{H^{\pm }}^2-M_{\Sigma }^2\right) \left(M_{H^{\pm }}^2+m^2-M_{\Sigma }^2\right)}\nonumber\\
&+\frac{2 M_{\Sigma }
   \left(m+M_{\Sigma }\right) C_0\left(0,0,4 m^2;M_{\Sigma }^2,M_{\Sigma }^2,M_{\Sigma }^2\right)}{-M_{H^{\pm }}^2-m^2+M_{\Sigma }^2}
\bigg]\nonumber\\
&+
\frac{\alpha_{em}  m^2 \sin (\alpha ) \cos (\alpha ) Y_N^{\alpha } Y_{\Sigma }^{\alpha }}{\pi }
\bigg[
-\frac{m_{\eta }^2 C_0\left(0,-m^2,m^2;m_{\eta }^2,m_{\eta }^2,m_{e_i}^2\right)}{m_{\eta }^2-m_{e_i}^2}\nonumber\\
&+ \frac{m_{e_i}^2 \left(m_{e_i}^2+m^2-m_{\eta }^2\right) C_0\left(0,-m^2,m^2;m_{e_i}^2,m_{e_i}^2,m_{\eta
   }^2\right)}{\left(m_{\eta }^2-m_{e_i}^2\right) \left(-m_{e_i}^2+m^2+m_{\eta }^2\right)}+\frac{2 m_{e_i}^2 C_0\left(0,0,4 m^2;m_{e_i}^2,m_{e_i}^2,m_{e_i}^2\right)}{-m_{e_i}^2+m^2+m_{\eta }^2}
\bigg]\nonumber\\
&+
\frac{\alpha_{em}  m^2 \cos ^2(\alpha ) (Y_{\Sigma }^{\alpha })^2}{2\sqrt{2} \pi }
\bigg[
\frac{m_{\eta }^2 C_0\left(0,-m^2,m^2;m_{\eta }^2,m_{\eta }^2,m_{e_i}^2\right)}{m_{\eta }^2-m_{e_i}^2}\nonumber\\
&-\frac{m_{e_i}^2 \left(m_{e_i}^2+m^2-m_{\eta }^2\right) C_0\left(0,-m^2,m^2;m_{e_i}^2,m_{e_i}^2,m_{\eta
   }^2\right)}{\left(m_{\eta }^2-m_{e_i}^2\right) \left(-m_{e_i}^2+m^2+m_{\eta }^2\right)}-\frac{2 m_{e_i}^2 C_0\left(0,0,4 m^2;m_{e_i}^2,m_{e_i}^2,m_{e_i}^2\right)}{-m_{e_i}^2+m^2+m_{\eta }^2}
\bigg]\nonumber\\
&+
\frac{\sqrt{2} \alpha_{em}  m^2 \sin ^2(\alpha ) (Y_N^{\alpha })^2}{2\pi }
\bigg[
\frac{m_{\eta }^2 C_0\left(0,-m^2,m^2;m_{\eta }^2,m_{\eta }^2,m_{e_i}^2\right)}{m_{\eta }^2-m_{e_i}^2}\nonumber\\
&-\frac{m_{e_i}^2 \left(m_{e_i}^2+m^2-m_{\eta }^2\right) C_0\left(0,-m^2,m^2;m_{e_i}^2,m_{e_i}^2,m_{\eta
   }^2\right)}{\left(m_{\eta }^2-m_{e_i}^2\right) \left(-m_{e_i}^2+m^2+m_{\eta }^2\right)}-\frac{2 m_{e_i}^2 C_0\left(0,0,4 m^2;m_{e_i}^2,m_{e_i}^2,m_{e_i}^2\right)}{-m_{e_i}^2+m^2+m_{\eta }^2}
\bigg]\nonumber\\
&-
\frac{8 \sqrt{2} \alpha_{em}  m^2 \cos ^2(\alpha ) M_W^2}{\pi  \left(M_{\Sigma }^2-M_W^2\right) \left(4 v_{\Omega }^2+v_{\phi }^2\right) \left(m^2-M_{\Sigma }^2+M_W^2\right) \left(m^2+M_{\Sigma }^2-M_W^2\right)}\nonumber\\
&
\bigg[
4 \left(m^2-M_W^2\right) \left(M_{\Sigma }^2-M_W^2\right) \left(m^2-M_{\Sigma }^2+M_W^2\right) C_0\left(0,0,4 m^2;M_W^2,M_W^2,M_W^2\right)\nonumber\\
&+2 M_{\Sigma } \left(2 m-M_{\Sigma }\right) \left(M_{\Sigma }^2-M_W^2\right)
   \left(m^2+M_{\Sigma }^2-M_W^2\right) C_0\left(0,0,4 m^2;M_{\Sigma }^2,M_{\Sigma }^2,M_{\Sigma }^2\right)\nonumber\\
   &-\left(m^2-M_{\Sigma }^2+M_W^2\right) \left(-M_W^2 \left(m^2+M_{\Sigma }^2\right)-4 m M_{\Sigma }
   \left(m^2+M_{\Sigma }^2-M_W^2\right)+4 M_{\Sigma }^4+M_W^4\right)\nonumber\\
   & C_0\left(0,-m^2,m^2;M_W^2,M_W^2,M_{\Sigma }^2\right)-M_{\Sigma } \left(m^2+M_{\Sigma }^2-M_W^2\right) \left(4 m^3-3 m^2 M_{\Sigma }+M_{\Sigma
   }^3-M_{\Sigma } M_W^2\right)\nonumber\\
   & C_0\left(0,-m^2,m^2;M_{\Sigma }^2,M_{\Sigma }^2,M_W^2\right)
\bigg]\,,
\end{align}
where, $C_0$ is the Passarino–Veltman function~\cite{Passarino:1978jh}, $m=m_{\chi_1^0}$ is the DM mass, $m_{e_i}$ are the lepton masses of the SM,
$M_{H^{\pm}}$ is the mass of the new charged scalar of this model, $M_W$ is the $W$ gauge boson mass and $\alpha_{em}$ is the fine structure constant.

\subsection{Pure singlet DM (Scotogenic limit)}
\label{sec:pure-singlet-sigmagg}

The $\mathcal{B}$ factor in this case can be obtained from Eq.~\eqref{eq:B} taken $\alpha =\pi/2$, $Y_{\Omega}=0$. In this limit we have
\begin{align}
\mathcal{B} &=
\frac{\alpha_{em}  m^2 (Y_N^{\alpha })^2}{\sqrt{2} \pi }\bigg[
\frac{m_{\eta }^2 C_0\left(0,-m^2,m^2;m_{\eta }^2,m_{\eta }^2,m_{e_i}^2\right)}{m_{\eta }^2-m_{e_i}^2}\nonumber\\
&-\frac{m_{e_i}^2 \left(m_{e_i}^2+m^2-m_{\eta }^2\right) C_0\left(0,-m^2,m^2;m_{e_i}^2,m_{e_i}^2,m_{\eta
   }^2\right)}{\left(m_{\eta }^2-m_{e_i}^2\right) \left(-m_{e_i}^2+m^2+m_{\eta }^2\right)}-\frac{2 m_{e_i}^2 C_0\left(0,0,4 m^2;m_{e_i}^2,m_{e_i}^2,m_{e_i}^2\right)}{-m_{e_i}^2+m^2+m_{\eta }^2}\bigg]\,.
\end{align}
In the limit of $m\gg m_{e_i}$, this expression gives (we used \texttt{PackageX}~\cite{Patel:2015tea})
\begin{align}
\mathcal{B} &=
\dfrac{\alpha_{em} (Y_N^{\alpha })^2}{2\sqrt{2}\pi}
\left[\text{Li}_2\left(\dfrac{m^2}{m_{\eta}^2}\right)-\text{Li}_2\left(-\dfrac{m^2}{m_{\eta}^2}\right)\right]\,.
\end{align}
Therefore, using Eq.~\eqref{eq:sigmav-gg}, we have
\begin{align}
\sigma v = \dfrac{|\mathcal{B}|^2}{32\pi m^2}=\dfrac{\alpha_{em}^2(Y_N^{\alpha })^4}{256\pi^3}\left[\text{Li}_2\left(\dfrac{m^2}{m_{\eta}^2}\right)-\text{Li}_2\left(-\dfrac{m^2}{m_{\eta}^2}\right)\right]^2\,,
\end{align}
in agreement with Ref.~\cite{Garny:2015wea}.

\subsection{Pure triplet DM (Minimal DM limit)}
\label{sec:pure-triplet-sigmagg}

The $\mathcal{B}$ factor in this case can be obtained from Eq.~\eqref{eq:B} taken $Y_{\Omega}=Y_{N}^{\alpha}=Y_{\Sigma}^{\alpha}=v_{\Omega}=0$, $\alpha =0$ and $m=M_{\Sigma}$. In this limit we have
\begin{align}
\mathcal{B} &=
\frac{8 \sqrt{2} \alpha_{em}^2 m^2}{\sin(\theta_W)^2}
\bigg[
\frac{m^2 \left(M_W^2-2 m^2\right) C_0\left(0,-m^2,m^2;m^2,m^2,M_W^2\right)}{M_W^2 \left(M_W^2-m^2\right)}-\frac{2 m^2 C_0\left(0,0,4 m^2;m^2,m^2,m^2\right)}{M_W^2}\nonumber\\
&-\frac{4 \left(M_W^2-m^2\right) C_0\left(0,0,4
   m^2;M_W^2,M_W^2,M_W^2\right)}{M_W^2-2 m^2}\nonumber\\
   &+\frac{\left(-4 m^4+2 m^2 M_W^2+M_W^4\right) C_0\left(0,-m^2,m^2;M_W^2,M_W^2,m^2\right)}{2 m^4-3 m^2 M_W^2+M_W^4}
\bigg]\,.
\end{align}
In the limit of $m\gg M_W$ we get (we used \texttt{PackageX}~\cite{Patel:2015tea}) 
\begin{align}
\mathcal{B} 
&\approx \frac{\sqrt{2} \pi  \alpha_{em}^2 m \left(-8 m^2+\pi  m M_W+4 M_W^2\right) \csc ^2\left(\theta _W\right)}{M_W^3-m^2 M_W}\,,
\end{align}
and therefore, using Eq.~\eqref{eq:sigmav-gg}, we get 
\begin{align}
\sigma v = \dfrac{|\mathcal{B}|^2}{32\pi m^2}\approx
\frac{4 \pi  \alpha_{em}^4 }{M_W^2 \sin(\theta_W)^4}\,,
\end{align}
in agreement with Ref.~\cite{Cirelli:2005uq}.

\bibliographystyle{jhep}
\bibliography{references}

\end{document}